\documentclass[prd,aps,showpacs,preprintnumbers,amsmath,amssymb,nofootinbib,preprintnumbers]{revtex4}
\usepackage{graphicx}
\usepackage{epsf}
\usepackage{amsmath}
\usepackage{amssymb,latexsym}
\usepackage{epstopdf}
\usepackage{bm}
\usepackage{color}
\usepackage{tabularx}
\usepackage{enumitem}
\usepackage{float}
\usepackage{array,booktabs}
\usepackage{footnote}
\usepackage{threeparttable}
\usepackage{graphicx}
\usepackage{hyperref}
\usepackage{amssymb,epsf}
\usepackage{latexsym}
\usepackage{epstopdf}
\usepackage{epsfig}
\usepackage{subfigure}

\begin{document}
     \title{Gauss-Bonnet black holes in a special anisotropic scaling spacetime}
     \author{
        S. Mahmoudi$^{1,2}$\footnote{email address: S.mahmoudi@shirazu.ac.ir},
        Kh. Jafarzade$^{1,2}$\footnote{email address: khadije.jafarzade@gmail.com} and
        S. H. Hendi$^{1,2,3}$\footnote{email address: hendi@shirazu.ac.ir}}
     \affiliation{
        $^1$Department of Physics, School of Science, Shiraz University, Shiraz 71454, Iran \\
        $^2$Biruni Observatory, School of Science, Shiraz University, Shiraz 71454, Iran \\
        $^3$Canadian Quantum Research Center 204-3002 32 Ave Vernon, BC V1T 2L7 Canada}

\begin{abstract}
Inspired by the Lifshitz gravity as a theory with anisotropic
scaling behavior, we suggest a new $(n+1)-$dimensional metric in
which the time and spatial coordinates scale anisotropically as
$(t,r,\theta_{i})\,\to
(\lambda^{z}t,\lambda^{-1}r,\lambda^{x_i}\,\theta_{i})$. Due to
the anisotropic scaling dimension of the spatial coordinates, this
spacetime does not support the full Schr\"{o}dinger symmetry
group. We look for the analytical solution of Gauss-Bonnet gravity
in the context of the mentioned geometry. We show that
Gauss-Bonnet gravity admits an analytical solution provided that
the constants of the theory are properly adjusted. We obtain an
exact vacuum solution, independent of the value of the dynamical
exponent $z$, which is a black hole solution for the
pseudo-hyperbolic horizon structure and a naked singularity for
the pseudo-spherical boundary. We also obtain another exact
solution of Gauss-Bonnet gravity under certain conditions. After
investigating some geometrical properties of the obtained
solutions, we consider the thermodynamic properties of these
topological black holes and study the stability of the obtained
solutions for each geometrical structure.
\end{abstract}

\maketitle
\section{Introduction}

Recently, there has been considerable interest in studying the AdS/CFT
correspondence which describes the duality between weakly coupled classical
relativistic gravitational systems in anti-de Sitter (AdS) spacetime and
certain strongly coupled conformal field theory (CFT) \cite{Witten:1998qj,
Gubser:1998bc, Maldacena:1997re}. This correspondence provides a new tool to
analyze the systems in some branches of physics such as condensed matter
physics \cite{Hartnoll:2009sz}, QCD quark-gluon plasmas \cite{Kovtun:2004de}
and atomic physics \cite{Maldacena:2008wh, Herzog:2008wg}. Over the last
several years, the framework of gravity-gauge duality has been generalized
to a much wider context than its original formulation and has been extended
beyond the relativistic domain. There are a variety of motivations behind
this generalization, a number of which are mentioned in what follows. Some
real condensed matter systems are described in the vicinity of their
critical temperatures by the non-relativistic conformal field theories since
they do not have relativistic symmetry in the proximity of this critical
point \cite{Hartnoll:2009sz, Herzog:2009xv}. Another application is using a
tool to cold atomic systems in the unitarity limit where the system of
two-component fermions interacting through a short-ranged potential which is
fine-tuned to support a zero-energy bound state and exhibits a
non-relativistic conformal symmetry in the limit of zero-range potential
\cite{Mehen:1999nd}. Besides, investigation of the high-temperature
superconductors in modern condensed matter physics is one of the targets of
gravity-gauge generalization \cite{Herzog:2009xv}. These attractive topics
motivate us to study the non-relativistic version of AdS/CFT duality in
order to gain more information about these systems. According to this
version of duality, spacetimes with the symmetry group known as Schr\"{o}%
dinger group are the gravity makes dual for the non-relativistic
scale-invariant systems enjoying Galilean symmetry \cite{Son:2008ye,
Balasubramanian:2008dm}. In this way, Lifshitz spacetimes, which were
originally introduced in \cite{Kachru:2008yh}, are understood as
gravitational theories dual to non-relativistic field theories at zero
temperature. This kind of gravity model exhibits scaling properties which
are anisotropic between space and time directions, i.e. $(t,x)\,\to
(\lambda^{z}t,\lambda\,x)$ . Such an anisotropic scaling has an effective
role in quantum phase transitions in condensed matter systems and ultracold
atomic gases \cite{Cardy:2002} and is characterized by the "dynamical
critical exponent" $z$ (where $z $ measures the degree of anisotropy between
space and time directions). Theories with $z\neq\,1$ are invariant under
non-relativistic transformations \cite{Kachru:2008yh, Taylor:2008tg} while
for $z=1$, the metric reduces to the relativistic isotropic scale invariance
spacetime corresponding to the AdS geometry.

It is worth mentioning that a metric with Lifshitz geometry is not a
solution to the vacuum Einstein field equations with a cosmological constant
except for $z =1$. For $z\neq\,1$, Einstein's gravity should be coupled with
other fields or modified by considering higher curvature terms investigated
in Refs. \cite{Dehghani:2010kd, Dehghani:2010gn, Brenna:2011gp,
Ghanaatian:2014bpa, Maeda:2011jj, Lee:2010iu, Alvarez:2014pra, Taylor:2008tg}%
. In other words, since there is nothing to produce an anisotropy in the
spacetime in pure Einstein gravity, Einstein's equations do not allow
anisotropic solutions. However, adding some higher-curvature tensors or
matter sources to this gravity may lead to anisotropic solutions for field
equations.

A special candidate for the higher curvature corrections to Einstein's
gravity is the Lovelock theory, which is the most general (purely metric)
gravitational theory leading to generally covariant field equations in
higher dimensions \cite{Lovelock:1971}. It is worthwhile to note that the
higher curvature gravity is also naturally obtained as next-to-leading term
in heterotic string effective action in the low-energy limit \cite%
{stringth1, stringth2, stringth3, stringth4, stringth5}. The
simplest Lovelock theory of gravity is well-known as Gauss-Bonnet
gravity including Einstein-Hilbert action, cosmological constant,
and quadratic curvature terms that is a topological invariant in
four and lower dimensions and is not effective in dynamics of the
system \cite{GB1, Lovelock:1971}. These curvature-squared terms
result in the significant feature that only up to the second-order
derivations of metric functions appear in the corresponding field
equations \cite{Boulware:1985wk, Zumino:1985dp, GB3, GB4}.
Moreover, this theory of gravity is a higher derivative gravity
enjoying the absence of ghost modes \cite{ghost1, ghost2}. It is worth mentioning that string theory predicts
some additional scalar fields coupled to the Gauss-Bonnet
invariant that are important in the appearance of non-singular
early time cosmologies and the late-time cosmic acceleration
\cite{Nojiri:2005vv, Nojiri:2010wj}. Strictly speaking,
considering the heterotic string effective action (loop corrected
superstring effective action), one can find the contribution of
the Gauss-Bonnet term as well as the existence of a dilatonic
field. However, by requiring the dilaton being a constant at the
Lagrangian level, one can obtain a pure Gauss-Bonnet term without
a scalar field. Nevertheless, there are some important motivations
to investigate Gauss-Bonnet gravity without additional
scalar/vector fields. For instance, almost the complete study of
dilatonic Gauss-Bonnet gravity requires numerical or perturbative
evaluation, while one can obtain analytical solutions in the
absence of a dilaton field. Besides, from the AdS/CFT
correspondence viewpoint, the Gauss-Bonnet term can be viewed as
the next-to-leading order corrections of large N expansion of
boundary CFTs in the strong coupling limit \cite{Nojiri:2000gv}.
As a result, various aspects of Gauss-Bonnet gravity and its
thermodynamic properties have been addressed in literature
\cite{blackGB1,blackGB2, blackGB3, blackGB4, blackGB5}. In the
context of holography, the effects of the Gauss-Bonnet gravity on
different properties of the system have been explored in some
physical contexts such as finite coupling \cite{holog1},
second-order transport \cite{holog2}, entanglement
entropy \cite{holog31, holog32, holog33, holog34} and superconductivity \cite%
{holog4, holog5, holog6}.

 Here, we consider Gauss-Bonnet gravity in a modified
Lifshitz geometry background. To generalize the Lifshitz gravity,
one could engineer theories that do not admit Galilean boosts and
then experience the anisotropic scaling spatial coordinates. These
theories have some applications in condensed matter systems,
including optimally doped cuprates and nonfermi liquid metals near
heavy electron critical points \cite{Con1,Con2,Con3,Con4}.
Besides, there is more evidence that such anisotropies exist in
our universe. For instance, it was investigated in Ref.
\cite{Mann:1a} that due to the discrete structure of the
spacetime, the first-order quantum corrections may lead to space
anisotropy. Such spatial anisotropies will have measurable
consequences at short distances if the corrections are between the
electroweak and the Planck scale. The mentioned corrections can be
incorporated using an anisotropic generalized uncertainty
principle (GUP), where the deformation from quantum gravity
depends on the direction chosen. One of the main motivations to
study the anisotropic GUP is to explain the
observed Cosmic Microwave Background (CMB) anisotropies \cite%
{Kiefer:1a,Bini:2a}. Investigation of the CMB anisotropies has a significant
role in developing the modern cosmology and our understanding of the very
early universe \cite{Bucher:2ab}. Another system which is spatially
anisotropic is super Yang-Mills (SYM) plasma \cite{Mateos:2ab}. In fact, the
plasma created in a heavy-ion collision can be locally anisotropic for some
short time after the collision, $T < T_{iso}$, and then becomes locally
isotropic \cite{Florkowski:2ab}. The investigation of anisotropic
Lifshitz-like geometries may contribute to the application of holographic
methods to this type of system.

Here, motivated by the mentioned observation of anisotropic scaling
behaviors, we propose a geometry in which the spatial coordinates also scale
anisotropically. We try to investigate the existence of this class of
geometry as a solution in Gauss-Bonnet gravity in the vacuum. Since the
higher power curvature terms seem to play the role of the desired matter
field, we check whether Gauss-Bonnet gravity can support the proposed
geometry in the vacuum. Our calculations show that this demand is met under
certain conditions. We also explore the properties of the black hole/brane
solution in Gauss-Bonnet gravity.

This paper is organized as follows: After the introduction, we briefly
review theories with an anisotropic scaling between time and space in
Section \ref{two}, and then we give a particular anisotropic metric where
the spatial coordinates scale anisotropically as well. In Sec. \ref{three},
we consider $n+1-$dimensional Gauss-Bonnet action and obtain field equations
under the mentioned metric ansatz. With these equations in hand, we solve
the field equations analytically in vacuum and discuss the main properties
of the solution in \ref{minus} and \ref{plus}. Thermodynamic behavior and
the stability of the black hole solutions are investigated in \ref{minusA}
and \ref{plusA}. Finally, we end the paper with remarkable results in Sec. %
\ref{five}.


\section{Background with anisotropic scaling dimensions}

\label{two}

Taking the problem of renormalizability in UV scale into account, one may
consider a Lorentz violation theory in which the higher spatial derivatives
are decomposed from higher time one. In this regard, the Horava-Lifshitz
approach is a method that assists us in constructing a theory with an
anisotropic scaling between time and space as
\begin{equation}
t\rightarrow \lambda ^{z}t,\hspace{1cm}\mathbf{r}\rightarrow \lambda\,
\mathbf{r},  \label{scaling1}
\end{equation}
where $z$ is the dynamical critical exponent \cite{Horava:2009uw}.
Generally, a family of backgrounds that geometrically follows this type of
scaling symmetry is known as Lifshitz spacetimes, which are a generalization
of AdS space and in arbitrary $(n+1)-$dimensions can be expressed as
\begin{equation}
ds^2 =-\frac{r^{2z}}{l^{2z}}dt^{2}+\frac{l^2 dr^{2}}{r^{2}} + r^2
\sum_{i=1}^{n-1}\,d\theta_{i}^{2},
\end{equation}
with the following scale transformation
\begin{equation}
t \to \lambda^{z}t,~~~~~~~~~~~\theta_{i} \to \lambda \theta_{i},~~~~~~~~~~r
\to {\lambda^{-1}}r~,
\end{equation}
where $z=1$ gives the usual metric on $AdS_{n+1}$. However, we can consider
a situation in which the spatial coordinates also scale anisotropically. To
this end, we suggest the following form of the metric
\begin{equation}  \label{metric2}
ds^2 =-\frac{r^{2z}}{l^{2z}}dt^{2}+\frac{l^2 dr^{2}}{r^{2}} +l^2
\sum_{i=1}^{n-1}\,\frac{r^{2x_{i}}}{l^{2x_{i}}}\,d\theta_{i}^{2},
\end{equation}
which is invariant under the following scale transformations
\begin{equation}
t \to \lambda^{z}t,~~~~~~~~~~~\theta_{i} \to \lambda^{x_{i}}
\theta_{i},~~~~~~~~~~r \to {\lambda^{-1}}r~,
\end{equation}
where $z$ and $x_{i}$ play the role of dynamical exponents.

More generally, the metric of an $(n+1)$-dimensional static spacetime with
different geometry that asymptotically (in a special limit) goes to the
metric \eqref{metric2} can be written as
\begin{equation}
ds^{2}=-\frac{r^{2z}}{l^{2z}}f(r)dt^{2}+\frac{l^2 dr^{2}}{r^{2}g(r)}
+l^{2}\,d\Omega ^{2}_{n-1,k},  \label{newmet3}
\end{equation}
where the functions $f(r)$ and $g(r)$ should go to $1$ asymptotically ($r
\rightarrow \infty$). Also, $d\Omega^{2}$ is the metric of an $(n-1)-$%
dimensional hypersurface which can be written as
\begin{equation}
d\Omega ^{2}_{n-1,k}=\left\{
\begin{array}{cc}
\frac{r^{2x_{{\tiny {1}}}}}{l^{2x_{1}}}\,d\theta
_{1}^{2}+\sum\limits_{i=2}^{n-1}\prod\limits_{j=1}^{i-1}\frac{r^{2x_{i}}}{%
l^{2x_{i}}}\sin ^{2}\theta _{j}d\theta _{i}^{2} & k=1 \\
\sum\limits_{i=1}^{n-1}\frac{r^{2x_{i}}}{l^{2x_{i}}}\,d\theta _{i}^{2} & k=0
\\
\frac{r^{2x_{1}}}{l^{2x_{1}}}\,d\theta _{1}^{2}+\,\sinh ^{2}\theta _{1}\left(%
\frac{r^{2x_{2}}}{l^{2x_{2}}}\,d\theta
_{2}^{2}+\sum\limits_{i=3}^{n-1}\prod\limits_{j=2}^{i-1}\frac{r^{2x_{i}}}{%
l^{2x_{i}}}\,\sin ^{2}\theta _{j}d\theta _{i}^{2}\right) & k=-1%
\end{array}
\right.
\end{equation}

For $(x_{i} ,z) > 1$, the scaling is asymmetric between the time and spatial
directions and the system is scale invariance without (along with) conformal
invariance for $k=\pm 1 (k=0)$.

It is important to note that Einstein's gravity with a negative cosmological
constant does not admit the metric \eqref{metric2} and at least a matter
source is required to engineer the scale invariant anisotropic background.
However, in what follows, we show that one may have the geometry %
\eqref{metric2} as an analytical solution of Gauss-Bonnet gravity without
matter, indicating that the higher curvature terms could have the desired
effect that matter fields induce.

Before ending this section, it should be mentioned that since we are looking
for exact solutions reducing to the metric \eqref{metric2} while $r$
goes to infinity, we always impose the following condition on the solutions
throughout the paper
\begin{equation}  \label{condition3}
\lim_{r \to \infty}\,f(r)\,=\,\lim_{r \to \infty}\,g(r)\,=\,1.
\end{equation}

\section{Black Hole/Brane Solutions in Gauss-Bonnet Gravity}\label{three}

The Einstein-Hilbert action with a Gauss-Bonnet term in the presence of
cosmological constant can be written down as \cite{Boulware:1985wk}
\begin{equation}
S=\frac{1}{16\pi G}\int d^{n+1}x\sqrt{-g}\left( R-2\Lambda \,+\alpha
\mathcal{L}_{GB}\right) ,  \label{3eq1}
\end{equation}%
where $\mathcal{L}_{GB}=R_{\mu \nu \gamma \delta }R^{\mu \nu \gamma \delta
}-4R_{\mu \nu }R^{\mu \nu }+R^{2}$ and $\alpha $ indicates the Gauss-Bonnet
coefficient with dimension $(length)^{2}$ which is positive according to the
phenomenological string theory \cite{Boulware:1985wk}. The gravitational field equation is
obtained by variation of the action with respect to the metric, resulting in
\begin{equation}
G_{\mu \nu }\,+\,\Lambda \,g_{\mu \nu }=2\alpha \left( \frac{1}{4}g_{\mu \nu
}\mathcal{L}_{GB}-RR_{\mu \nu }+2R_{\mu \gamma }R_{\ \nu }^{\gamma
}+2R_{\gamma \delta }R_{\ \mu \ \ \nu }^{\gamma \ \delta }-R_{\mu \gamma
\delta \lambda }R_{\nu }^{\ \gamma \delta \lambda }\right) .
\label{field eq}
\end{equation}%
According to what mentioned before, we consider the black hole/brane
solutions in an $(n+1)$-dimensional anisotropic spacetime. To this end, using
the metric \eqref{newmet3}, the field equation \eqref{field eq} reduces to
the following differential equations
\begin{equation}
E_{1}=2k\left[ \hat{\alpha}g^{\prime }(r)rx-\frac{(n-2)\Upsilon }{2}\right]
\left( {\frac{r}{l}}\right) ^{2x}+nx\bigg[rg^{\prime }(r)\frac{\Upsilon }{n}%
-x^{3}\hat{\alpha}g^{2}(r)+xg(r)l^{2}+\frac{2\Lambda l^{4}}{nx(n-1)}\bigg]({%
\frac{r}{l}})^{4x}-k^{2}\hat{\alpha}(n-4)=0,  \label{Eq1}
\end{equation}%
%
%
%
%
%
%
%
%
\begin{eqnarray}
E_{2} &=&\,k\bigg{[}x\hat{\alpha}g(r)f^{\prime }(r)r-\frac{f(r)}{2}\bigg({l}%
^{2}(n-2)-x\hat{\alpha}\bigg \{2\,x(n-4)+4z(n-3)(n-2)\bigg\}g(r)\bigg )%
\bigg{]}({\frac{r}{l}})^{2x}-\frac{1}{2}{k}^{2}\hat{\alpha}f(r)(n-4)+  \notag
\\
&&\left\{ \frac{xr\Upsilon g(r)f^{\prime }(r)}{2}+f(r)\bigg{[}\left( 1+\frac{%
{x}(n-2)}{2z}\right) xz{l}^{2}g(r)-{x}^{3}\hat{\alpha}g^{2}(r)\left( \frac{x%
}{2}(n-4)+2z\right) +\frac{\Lambda {l}^{4}}{(n-1)}\bigg{]}\right\} ({\frac{r%
}{l}})^{4x}=0,
\end{eqnarray}%
%
%
%
%
%
%
%
%
\begin{eqnarray}
E_{3} &=&\bigg{[}\frac{1}{2}g(r)f(r){r}^{2}\Upsilon \left( f^{\prime \prime
}(r)-\frac{1}{2}{f^{\prime }}^{2}(r)\right) +rf^{\prime }(r)f(r)\Bigg(\frac{%
rg^{\prime }(r)\left[ \Upsilon -4\hat{\alpha}g(r){x}^{2}\right] }{4}+\frac{%
\Upsilon g(r)\left[ (n-2)x+2z+1\right] }{2}\Bigg )  \notag \\
&&+rg^{\prime }(r)f^{2}(r)\Bigg(\left( \frac{\left[ (n-2)x+z\right] {l}^{2}}{%
2}-g(r){x}^{2}\hat{\alpha}\left[ (n-4)x+3z\right] \right) +{l}^{2}g(r)\Big\{%
\frac{(n-1)(n-2){x}^{2}}{2}+(n-2)xz+{z}^{2}\Big\}  \notag \\
&&-2g^{2}(r){x}^{2}\hat{\alpha}\Big\{\frac{(n-4)(n-1){x}^{2}}{4}+(n-2)xz+{z}%
^{2}\Big\}+\Lambda {l}^{4}\Bigg )\bigg{]}({\frac{r}{l}})^{2x}-\frac{1}{2}\,{k%
}^{2}\hat{\alpha}f^{2}(r)(n-4)(n-5)({\frac{r}{l}})^{-2x}  \notag \\
&&+k\hat{\alpha}g(r)f(r)f^{\prime \prime }(r){r}^{2}-\frac{1}{2}k\hat{\alpha}%
g(r)(f^{\prime }(r))^{2}{r}^{2}+k\hat{\alpha}rf(r)f^{\prime }(r)\bigg(\frac{1%
}{2}rg^{\prime }(r)+g(r)\Big \{(n-4)x+2z+1\Big\}\bigg )  \notag \\
&&-krg^{\prime }(r)f^{2}(r)\Big[\frac{(n-2)(n-3){l}^{2}}{2}-\hat{\alpha}%
\Big\{(n-4)x+z\Big\}-\hat{\alpha}(n-4)g(r)\Big\{(n-3){x}^{2}+2xz+\frac{2{z}%
^{2}}{n-4}\Big\}\Big]=0,  \label{Eq3}
\end{eqnarray}%
where $\Upsilon ={l}^{2}-2\hat{\alpha}g(r){x}^{2}$, prime denotes the
derivative with respect to $r$ and we have defined $\hat{\alpha}\equiv
(n-2)(n-3)\alpha $ for the sake of brevity. It should be mentioned that in
order to have a consistent solution, we have to set $x_{i}=x$. For
convenience, in the next section we have applied this condition in the
metric from the beginning.

\subsection{Vacuum Solutions}

\label{three1}

We first investigate the possibility of having an $(n+1)$-dimensional
solution of our proposed metric 
in the absence of matter field. To this end, one can easily find that for
arbitrary $z$, the following metric function can be obtained
\begin{equation}
g(r)=1+k\frac{\hat{\alpha}_{\text{{\tiny {eff}}}}}{r^{2x}},  \label{gvac}
\end{equation}
where
\begin{equation}  \label{alphaeff}
\hat{\alpha}_{\text{{\tiny {eff}}}}=\,\frac{\tilde{\alpha}^{x}\pm\big(\tilde{%
\alpha}^{2x}-x^{3}\tilde{\alpha}^{2}\big)^{1/2}\delta_{n,4}}{x^2},
\end{equation}
provided that the following constraints on the cosmological constant and
Gauss-Bonnet coefficient are applied
\begin{equation}
\Lambda =-\frac{n(n-1)x^{2}}{4l^{2}}\text{ \ \ and \ \ } \tilde{\alpha}=2x^2%
\hat{\alpha}=l^{2},  \label{Coef1}
\end{equation}
or equivalently $\Lambda =-\frac{n(n-1)}{8\hat{\alpha}}$. We should note
that, here, the value of the cosmological constant and Gauss-Bonnet
coefficient depend on the dynamical exponent $x$ which is the scaling
dimension of the $\theta_{i}$ coordinates.

According to \eqref{alphaeff} and dimensional analysis, one can easily find
that the dynamical exponent $x$ must be equal to one for $5-$dimensional
spacetime and hence, our proposed metric in $5-$dimension reduces to that in
Lifshitz gravity. However, in higher dimensions, there is no restriction on
the value of the parameter $x$. Therefore, in the following, we will
concentrate on the dimensions higher than $5$.

Notably, the function of $f(r)$ cannot be determined by the field equations.
In fact, substituting \eqref{gvac} into \eqref{Eq1}-\eqref{Eq3} causes the
field equations to become zero, independent of the function of $f(r)$. The
degeneracy of the field equations which leads to the arbitrary function of $%
f(r)$ has been previously considered in $5-$dimensional
Einstein-Gauss-Bonnet gravity with a cosmological constant \cite%
{Dotti:2007az}. However, this degeneracy may be removed in the presence of
the matter field \cite{Dehghani:2010kd}.

\subsection{General Solutions}

\label{three2} Generally, the solutions of our suggested metric %
\eqref{newmet3} in the Einstein-Gauss-Bonnet gravity will be obtained as
follows
\begin{equation}
g(r)=f(r)=1+k\frac{\tilde{\alpha}^{x}}{x^2r^{2x}}\pm \sqrt{\frac{16 \pi
G\hat \alpha M}{x(n-1) r^{nx}}},  \label{Msol}
\end{equation}
providing that the constraints \eqref{Coef1} are held and the dynamical
exponent $x$ and $z$ have the same values. Moreover, in order to correctly
reproduce the asymptotic behavior, we have imposed the condition %
\eqref{condition3} on the solution. Also, in the above relation,
$M$ is an integration constant known as the mass of the solution
per unit volume, i.e. $M=\frac{m}{\Sigma_k}$, and $G$ is the
gravitational constant. It should be mentioned that the
coefficient of the mass term is set by comparing the obtained
solution with the Gauss-Bonnet black holes in AdS spacetime
provided that $\hat\alpha=\frac{l^2}{4} $ \cite{GB4} and
considering the fact that the cosmological constant, here, for the
case of $x=1$ is half of that in the AdS spacetime. Also, note
that this solution has two branches with "$+ $" and "$- $" signs.
Before proceeding further, it is worth mentioning
two points: First, one may expect to obtain the usual AdS black
hole solutions in Gauss-Bonnet gravity for the case of $x=1$.
However, due to the constraints \eqref{Coef1}, our solution will
be equivalent to the AdS solution with
$\hat{\alpha}=-\frac{n(n-1)}{8\Lambda}$ with a difference that, as
noted previously, the cosmological constant for the case of $x=1$
in our model is half of that in AdS spacetime.
Second, since the null energy condition plays a
vitally important role in gravity \cite{Bardeen:1973gs,
Chatterjee:2012zh}, it is important to explore if the discussed
spacetime satisfies it. To this end, we use the  geometric form of
null energy condition \cite{Kontou:2020bta}
    \begin{equation}
         G_{\mu\nu}N^{\mu}N^{\nu} \geqslant 0, \nonumber
    \end{equation}
      where $N^{\mu}$  is a null vector. Choosing
      \begin{equation}
      N^{t}=(\frac{l}{r})^{x}\frac{1}{\sqrt{f(r)}},\,\quad\,N^{r}=\frac{r\sqrt{g(r)}}{l},\,\quad\,N^{i}=0,
      \end{equation}
      and making use of \eqref{field eq}, one can easily find that $G_{\mu\nu}N^{\mu}N^{\nu}=0$ for the solution \eqref{Msol} which means that
      independent of the dynamical exponent $x$, these spacetimes meet the null energy condition for three kinds of horizon topology, $k$.

In what follows, we explore different geometric and thermodynamic
properties of $\pm$ branches of the solution \eqref{Msol},
separately.

\section{Negative sign branch}\label{minus}

\begin{figure}[h!]
\centering\includegraphics[scale=0.4]{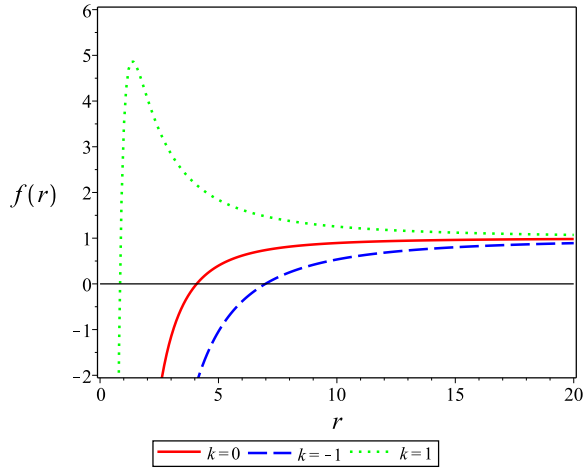}
\caption{$f(r)$ versus $r$ for $n=5$, $G=1$, $\protect\alpha=3$ , $M=5$ and $%
x=1$ for "$-$" sign branch.}
\label{fig1}
\end{figure}

To investigate some properties of the solution with "$-$" sign, we have
provided some diagrams related to the behavior of the metric function $f(r)$
in terms of $r$ for different model parameters. First, we have plotted the
function of $f(r)$ versus $r$ for $x = 1$ with $k = 0$ and $k = \pm 1$ in
Fig. \ref{fig1}. According to this figure, for $k=0,-1$ and $x=1$, the
metric functions increase from zero at $r = r_0$ to $1$ at $r =\infty$.
While, for $k = 1$ and $x = 1$, the metric function increases to larger than
unity at intermediate values of $r$ and then goes to $1$ at spatial infinity.

\begin{figure}[h!]
\centering
\subfigure[\, $k=0$]{\includegraphics[scale=0.25]{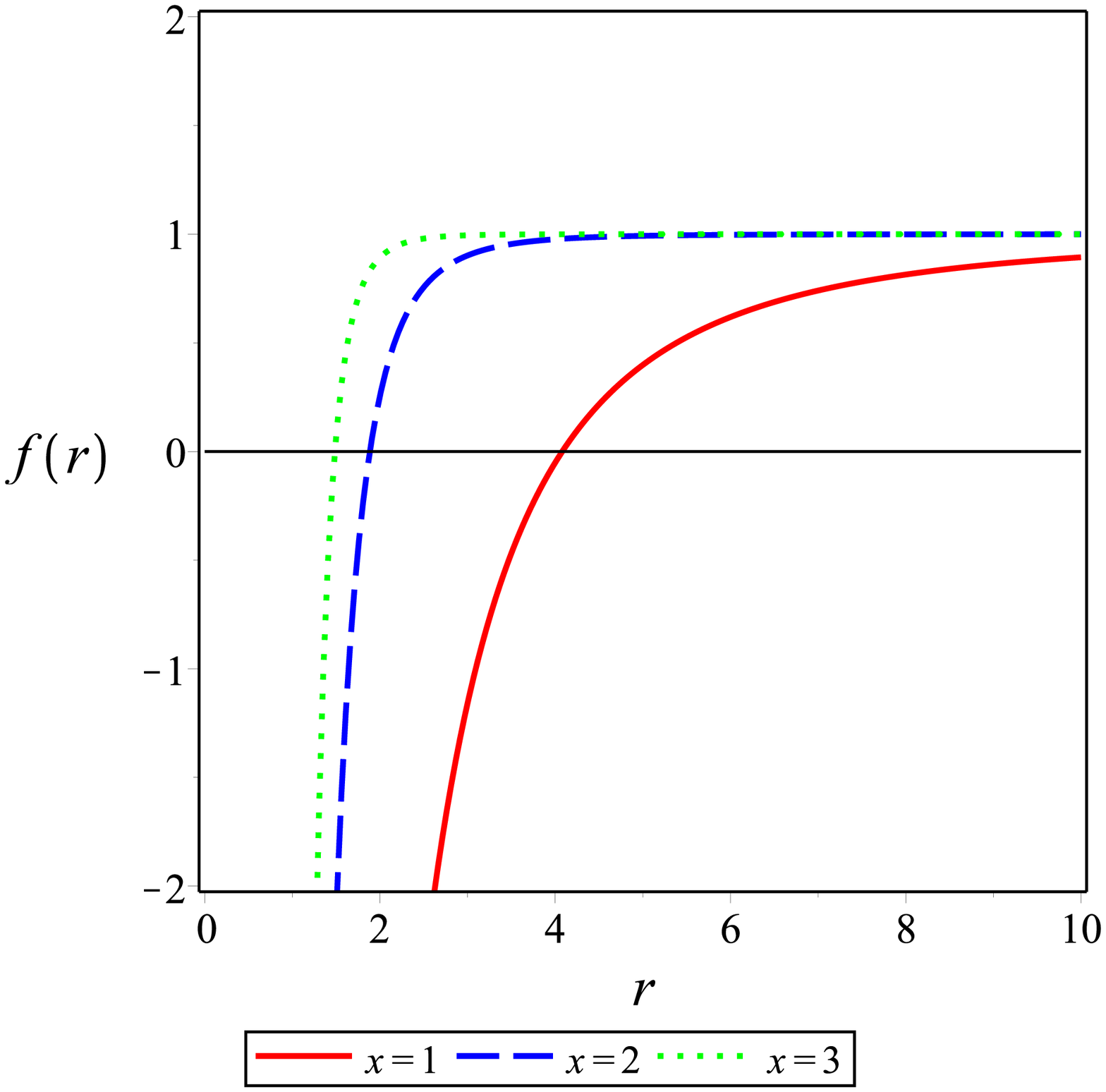}\label{fig2-1}}
\hspace*{.1cm}
\subfigure[\, $k=-1$
        ]{\includegraphics[scale=0.25]{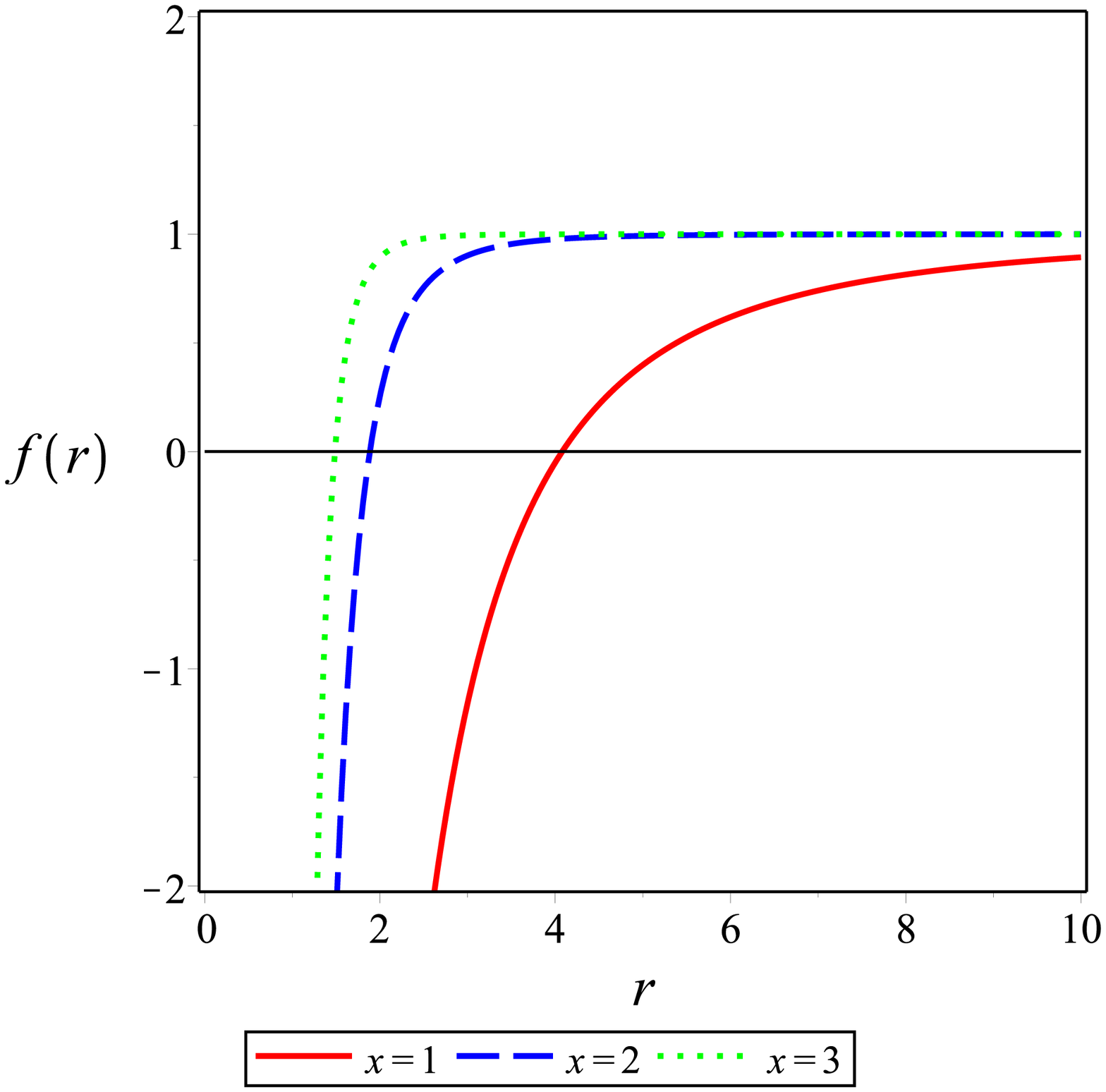}\label{fig2-2}}
\subfigure[\, $k=1$
        ]{\includegraphics[scale=0.25]{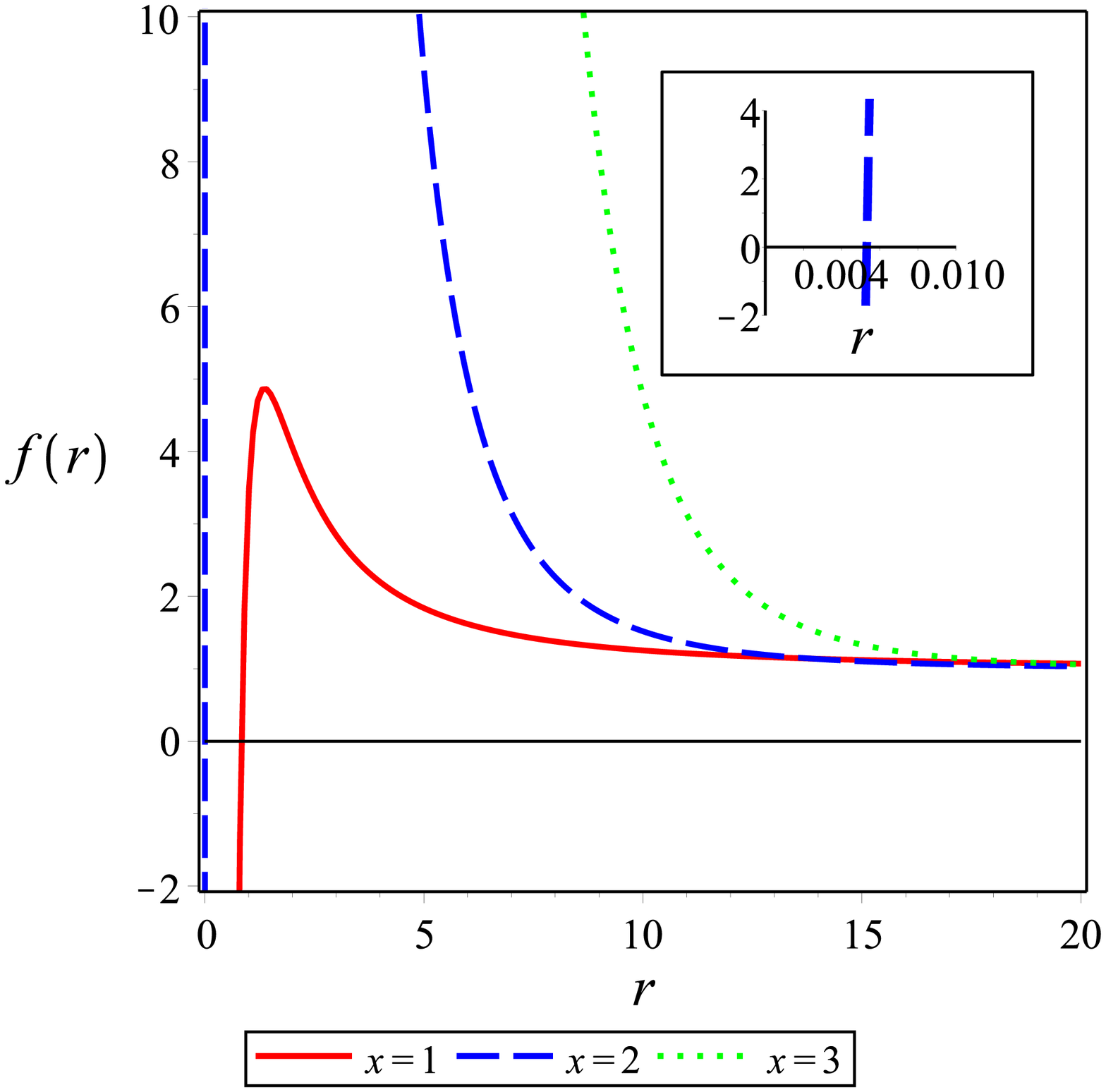}\label{fig2-3}}
\caption{Behavior of $f(r)$ versus $r$ for $n=5$, $G=1$, $\protect\alpha=3$
and $M=5$ for "$-$" sign branch.}
\label{fig2}
\end{figure}

Next, we try to study the effects of increasing the value of the dynamical
exponent $x$ on the solutions. In this regard, we have provided different
panels in Fig. \ref{fig2} that the behavior of the metric function in terms
of $r$ for each curvature structure is depicted for different values of $x$.
According to this figure, solutions related to $k=0$ and $k=-1$ are not very
sensitive to the parameter $x$. However, the solution with $k=1$ is very
sensitive to the change of $x$ and the metric function rapidly increases to
values larger than $1$ by changing the dynamical exponent $x$ from $1$ to
larger values.

Another important point is that the metric function with $"-"$ sign
interprets as a black hole with just one horizon for three geometric
structures of the horizon ($k=0\,,\pm 1$). To confirm our claim, one can
calculate $f(r)$ derivatives' roots, since the number of $f(r)$ derivatives'
roots shows the number of function's extrema. A straightforward calculation
shows that the derivative of the metric function with $k=0$ and $k=-1$ does
not have any real positive roots, independent of the metric parameters.
Therefore, these functions do not experience any extrema. However, the
derivative of the solution with $(k=1)$ has one root which is related to the
maximum point that is shown in Fig. \ref{fig1}. On the other hand, since
\begin{eqnarray}
&&\lim_{r \to \infty}\,f(r)\,\rightarrow\,1,  \notag \\
&&\lim_{r \to 0}\,f(r)\,\rightarrow\,-\infty,
\end{eqnarray}
the metric function $f(r)$ enjoys at least one root. According to the two
mentioned points, we conclude that the metric function with "$-$" sign is a
black hole that meets one horizon for all three geometric structures of the
horizon.

We now examine the singularity of the solutions by calculating the scalar
curvatures of the spacetime. To do so, we consider some curvature invariants
such as the Kretschmann scalar as well as Ricci scalar and investigate their
behavior in the presence of large and small radii. No matter the geometric
structures of the horizon, both Ricci and Kretschmann invariants diverge as $%
r \rightarrow 0$ 
\begin{equation}  \label{Kres}
\left. \mathcal{K}\right\vert _{{r\rightarrow 0}^{+}} = {\beta(n)\frac{GM\pi%
}{x\alpha r^{nx}}+\mathcal{O}(r^{-\frac{n+4}{2}x}) } \,\,\,\,\,\,\,\,\,\,\,%
\,\,\,\,\,\,\,\,\,n\geq 5,
\end{equation}
and
\begin{equation}  \label{Ricci}
\mathcal{R}\vert _{{r\rightarrow 0}^{+}} = \xi(n,k)\sqrt{ \frac{24GM\pi}{%
x\alpha}}r^{-\frac{nx}{2}}\,+\mathcal{O}(1)\,\,\,\,\,\,\,\,\,\,\,\,\,\,\,\,%
\,\,\,\,n\geq 5,
\end{equation}
where $\beta(n)$ and $\xi(n,k)$ are numbers which vary by changing
the dimensions of the spacetime and the value of $k$. Therefore,
this solution has an essential singularity at $r = 0$. Besides,
the mentioned scalars tend to
$\mathcal{R}=\,-\,\frac{n(n+1)}{2\hat{\alpha}}$ and
$\mathcal{K}=\,\frac{n(n+1)}{2\,\hat{\alpha}^2}$ when $r$ goes to
infinity, and thus, the spacetime has a constant curvature at
spatial infinity. It is worth mentioning that this is an expected
result since the line element (\ref{newmet3}) goes to the metric
\eqref{metric2} based on Eq. (\ref{condition3}).

\subsection{Black Hole/Brane Thermodynamics and Thermal Stability}\label{minusA}

In this section, we try to investigate the thermodynamic properties of this
branch of solution. According to Eq. \eqref{Msol}, the mass of black hole
per unit volume $\Sigma_{k}$ can be expressed in terms of the horizon radius
$r_+$ as
\begin{equation}  \label{Mass}
M=\frac{(n-1) r_+^{(n-2)x}}{8\pi G x}\frac{\tilde{\alpha}^{x}}{\hat{\alpha}} %
\Big[k +\frac{ k^2\tilde\alpha^x} {2x^2r_+^{2x}} +\frac{x^2r_+^{2x}}{%
2\,\tilde\alpha^{x}}\Big].
\end{equation}
It is straightforward to show that for $x=1$, Eq. (\ref{Mass}) reduces to
the mass of Gauss-Bonnet black holes in AdS spacetime (provided that $\hat{%
\alpha}=\frac{l^2}{4}$) \cite{GB4}, namely
\begin{equation}  \label{Mass1}
M=\frac{(n-1) r_{+}^{n-2}}{16\pi G}\Big[k +\frac{ k^2\hat\alpha} {r_+^{2}} +%
\frac{r_+^{2}}{l^2}\Big].
\end{equation}
It is also important to note that the difference in the coefficients of the
two relations originated from the difference between the cosmological constant
in our model and the Gauss-Bonnet black hole in the AdS spacetime.

To explore physical properties, we should determine temperature as the next
step. The Hawking temperature of the black holes can be calculated by the
surface gravity as
\begin{equation}  \label{temp}
T_H = \frac{1}{4\pi}\Big[\sqrt{-g^{tt}g^{rr}}~ g_{tt}^{\prime }(r)\Big]%
_{r=r_+}.
\end{equation}
For the branch with "$-$" sign, the temperature will be obtained as
\begin{eqnarray}
T_{{\tiny {H}}}={\frac {4\,{x}^{2}r_{+}^{2x}+ \left| {x}^{2}r_{+} ^{2x}+k
\tilde{\alpha} ^{x} \right|( n-4) }{8\pi\,x\, r_{+}^{x}\sqrt { \tilde{\alpha}
^{x+1}}}},  \label{Temp22}
\end{eqnarray}
which considering the points mentioned earlier about the difference between
the two models, one can easily check that the temperature for $x=1$ is in
agreement with that of the Gauss-Bonnet black hole in the AdS spacetime. In
the following, we study the situations in which the expression inside the
absolute value is positive. This condition requires that for the case of $%
k=-1$ the horizon radius is larger than $\left( {\frac { \tilde{\alpha} ^{x}%
}{{x}^{2}}} \right) ^{\frac{1}{2x}} $. Considering this condition, the
temperature relation \eqref{Temp22} reduces to the following relation
\begin{eqnarray}
T_{{\tiny {H}}}={\frac {r_{+}^{x}nx}{8\,\pi\,\sqrt { \tilde \alpha ^{x+1}}}
\left( 1+k\frac{n-4}{n}{\frac { \tilde\alpha ^{x}}{{x}^{2}r_{+}^{2x}}}
\right) }.  \label{Temp2}
\end{eqnarray}

Next, we try to compute the entropy of the black hole.
For this purpose, we use the fact that as a thermodynamic system, black
hole's entropy must obey the first law of black hole thermodynamics
\begin{equation}
\delta M = T \delta S.
\end{equation}
Integrating this relation along with considering the physical assumption
that the entropy will vanish if the horizon of the black hole shrinks to
zero results that
\begin{equation}  \label{entropy}
S =\int ^{r_{+}}_0 T^{-1}\left (\frac{\partial M} {\partial r_+}\right) dr_+.
\end{equation}
Hence, the entropy of the black hole per unit volume $\Sigma_{k}$ will be
obtained by substituting \eqref{Mass} and \eqref{Temp2} into \eqref{entropy}
as follows 
\begin{equation}  \label{entropy1}
S={\frac {r_{+}^{( n-1) x}{x}^{2}}{G}\sqrt {\tilde \alpha ^{(x-1)}} \left(
1+k\frac{n-1}{n-3}{\frac {\tilde{\alpha}^{x} }{ {x }^{2}r_{+}^{2x}}} \right)
}.
\end{equation}


Up to now, we showed that one can regard the obtained black hole as a
thermodynamic system. On the other hand, it is necessary to investigate the
stability of a thermodynamic system under thermal perturbations. In order to
investigate the local stability of black holes, we need to calculate the
heat capacity. It is known that the heat capacity of the black hole is
defined as
\begin{equation}  \label{heat capacity}
C=T\,\frac{\partial S}{\partial T},
\end{equation}
where for the obtained solution, it gets the following form via the chain
rule
\begin{equation}
C={\frac {{x}^{2}( n-1)r_+^{( n-1) x}\sqrt { \tilde \alpha ^{x-1}}}{G}
\left( 1+{\frac {k( n-4) \tilde\alpha^{x}}{n{x}^{2}r_+ ^{2\,x}}} \right)
\left( 1+{\frac {k \tilde\alpha ^{x}}{{x}^{2}r_+ ^{2x}}} \right) \left( 1-{%
\frac {k( n-4) \tilde \alpha ^{x}}{n{x}^{2}r_+ ^{2x}}} \right) ^{-1}}.
\end{equation}

Besides, global stability of the black holes can be explored with the help
of free energy of the system, defined as $F\,=\,M\,-\,T\,S$. Using %
\eqref{Mass}, \eqref{Temp2} and \eqref{entropy1}, one can obtain the
functional form of free energy with the following explicit relation
\begin{eqnarray}
F&=&-{\frac {r_+^{nx}x^3}{8 G\pi \tilde\alpha} \Bigg[ \left( 1+k{\frac {(
n-1)\tilde \alpha^{x}}{( n-3){x}^{2} r_+^{2x}}} \right) \left( 1+k{\frac {(
n-4) \tilde \alpha ^{x}}{n{x}^{2}r_+^{2x}}} \right) n- \left( 1+k{\frac {%
\tilde \alpha ^{x}}{{x}^{2}r_+^{2x}}} \right)^{2}(n-1) \Bigg] }.
\end{eqnarray}
So far, we got some thermodynamic quantities of Gauss-Bonnet black holes in
our proposed geometry. As can be seen, these quantities strongly depend on
the Gauss-Bonnet coefficient $\alpha$, critical exponent $x$, horizon
structure $k$, and spacetime dimensions $n$. In the following, we discuss
the stability of the solution in more details according to the
classification of the horizon structures, $k = 0$, $k = -1$ and $k = 1$,
respectively.

\vspace{0.5cm} ${\bullet{\mathbf{{\mathbf{Case\; k=0}: }}}} $ \vspace{0.2cm}

In the case of the solution with $k=0$, the thermodynamic quantities are
\begin{eqnarray}  \label{3eq15}
&& T= {\frac {nx{r_{+}^x}}{8\pi\sqrt { \tilde \alpha ^{x+1}}}} ,  \notag \\
&& S= {\frac {r_{+}^{(n-1) x}{x}^{2}\sqrt {\tilde \alpha ^{x-1}}}{G}} ,
\notag \\
&& C= {\frac {{x}^{2}( n-1) r_{+}^{( n-1) x}\sqrt { \tilde \alpha^{x-1}}}{G}}
,  \notag \\
&& F= -{\frac {r_{+}^{nx}x^3}{ 8 G\pi\tilde \alpha}} ,
\end{eqnarray}
where $r_+^{nx}=\frac{8\pi GM\tilde{\alpha}}{(n-1)x^2}$. Here, we note that
as the above relations show, these thermodynamic quantities are dependent on
the parameter $\tilde{\alpha}$ owing to the fact that the Einstein gravity
has no solution in the proposed geometry and the existence of the
Gauss-Bonnet term is necessary to have an appropriate exact solution.

Regarding the stability of the solution, one finds that it is quite locally
stable due to the positivity of the temperature and heat capacity for all
values of the model parameters. Moreover, the strictly decreasing behavior
of the free energy function guarantees the global stability of the solution.

\vspace{0.5cm} ${\bullet{\mathbf{{\mathbf{Case\; k=-1}: }}}} $ \vspace{0.2cm}

\begin{figure}[tbp]
\centering\includegraphics[scale=0.4]{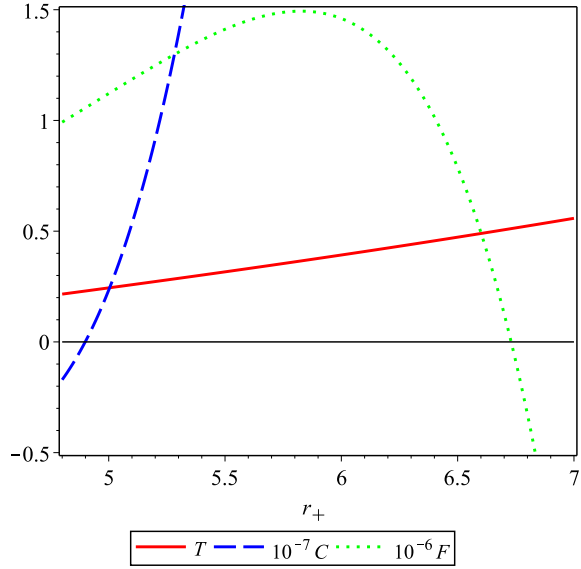}
\caption{Behavior of $T$, $C$ and $F$ versus $r_+$ for $k=-1$, $n=5$, $x=2$
and $\protect\alpha=1$ for "$-$" sign branch.}
\label{fig4}
\end{figure}

In this case, calculations show that the solutions are not always stable. To
be more clear, we have provided Fig. \ref{fig4} in which the behavior of
temperature, heat capacity, and free energy in terms of $r_+$ are depicted.
It is notable that we use different scales for temperature, heat capacity,
and free energy to make them comparable with each other.

The temperature of the black hole is always positive. However, the heat
capacity is negative for the following horizon radius interval
\begin{equation}
r_+\,\leqslant\, {r_{+}}_{{\tiny {min}}},
\end{equation}
where ${r_{+}}_{{\tiny {min}}}=\left( {\frac { \tilde\alpha^{x}}{{x}^{2}}}
\right) ^\frac{1}{2x}$ is the smallest horizon radius of the stable black
hole. The temperature for the smallest stable black hole is
\begin{equation}
T_{{\tiny {\text{min}}}}={\frac {1}{2\pi\sqrt {\tilde \alpha}}}.
\end{equation}
Hence, the temperature of the stable black holes starts from $T_{{\tiny {%
\text{min}}}}$ at ${r_{+}}_{{\tiny {\text{min}}}}$ and monotonically goes to
infinity as $r_{+}\rightarrow\infty$. The existence of a lower limit on the
horizon radius and also temperature would be interesting from the
information paradox point of view which may be investigated in the future.
Moreover, the free energy of the smallest stable black hole is obtained as
\begin{equation}
F_{min}={\frac {x^3}{\pi G ( n-3 ) \tilde{\alpha}} \left( \sqrt {{\frac {%
\tilde\alpha^{x}}{{x}^{2}}}} \right) ^{n}}.
\end{equation}

Our calculation shows that the free energy always experiences a positive
maximum value, $F_{{\tiny \text{max}}}={\frac {x^3}{2( n-1 )G\pi \tilde\alpha%
} \left( \sqrt {{\frac {( n-1) \tilde \alpha ^{x} }{( n-3 ){x}^{2} }}}
\right) ^{n}} $ at $r_+=\left( {\frac {( n-1) \tilde{\alpha} ^{x}}{ ( n-3 ){x%
}^{2} }}\right) ^{\frac{1}{2x}}$ and then goes to negative infinity as $r_+
\rightarrow\,\infty$. As an example, the behavior of this function in terms
of the horizon radius for a set of model parameters is plotted in Fig. \ref%
{fig4}.

\begin{figure}[h!]
\subfigure[ $10^{-4}F-r_+$(red), $6\times10^{-6}F-r_+$(blue) and
$7.7\times10^{-8}F-r_+$(green) for $n=5$, $x=2$ and
$G=1$.]{\includegraphics[scale=0.25]{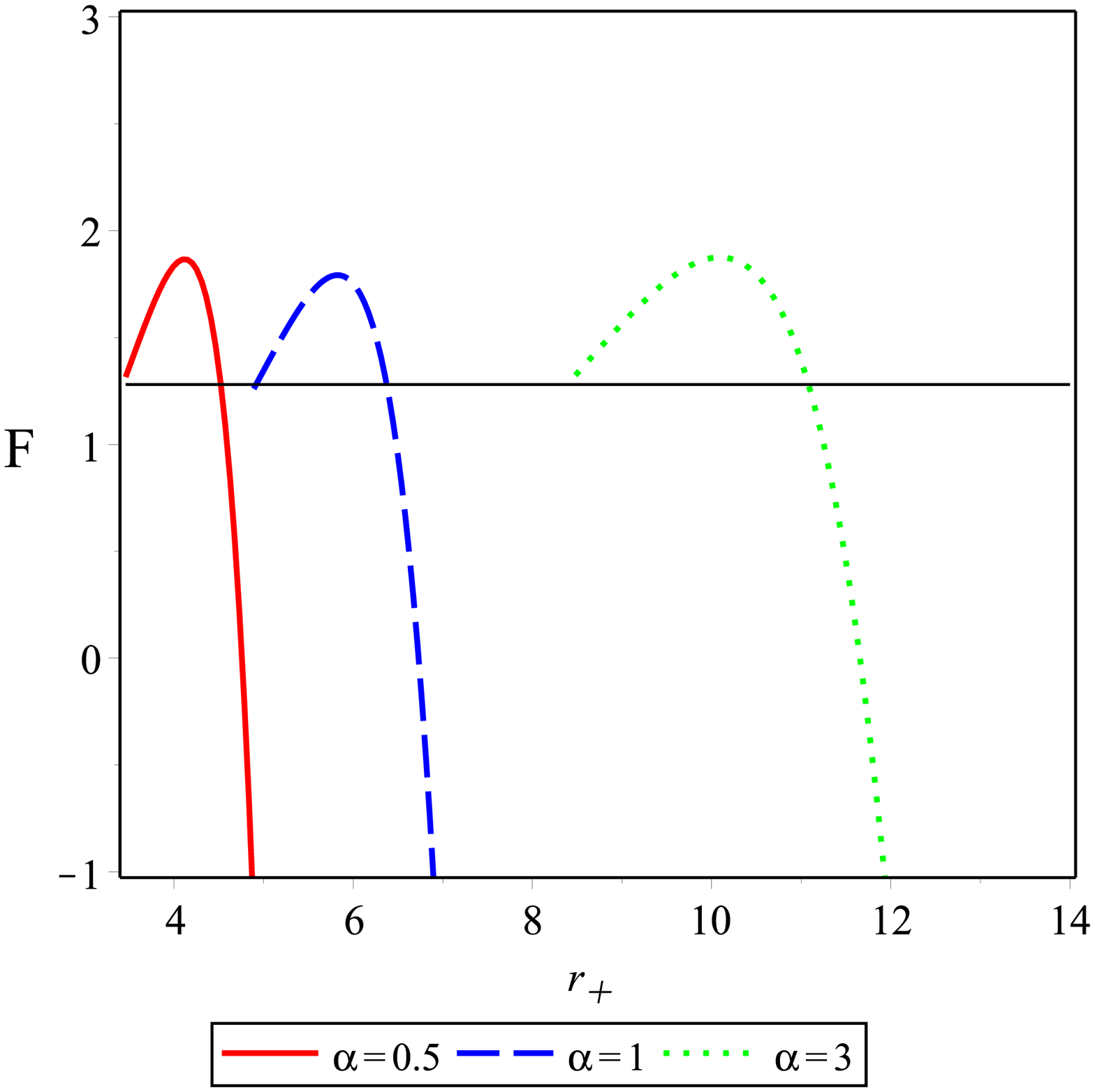}\label{fig51}} \hspace*{.5cm}
\subfigure[ $10^{-1}F-r_+$(red), $10^{-5}F-r_+$(blue) and $7\times10^{-12}F-r_+$(green) for $n=5$, $\alpha=1$ and $G=1$.
    ]{\includegraphics[scale=0.25]{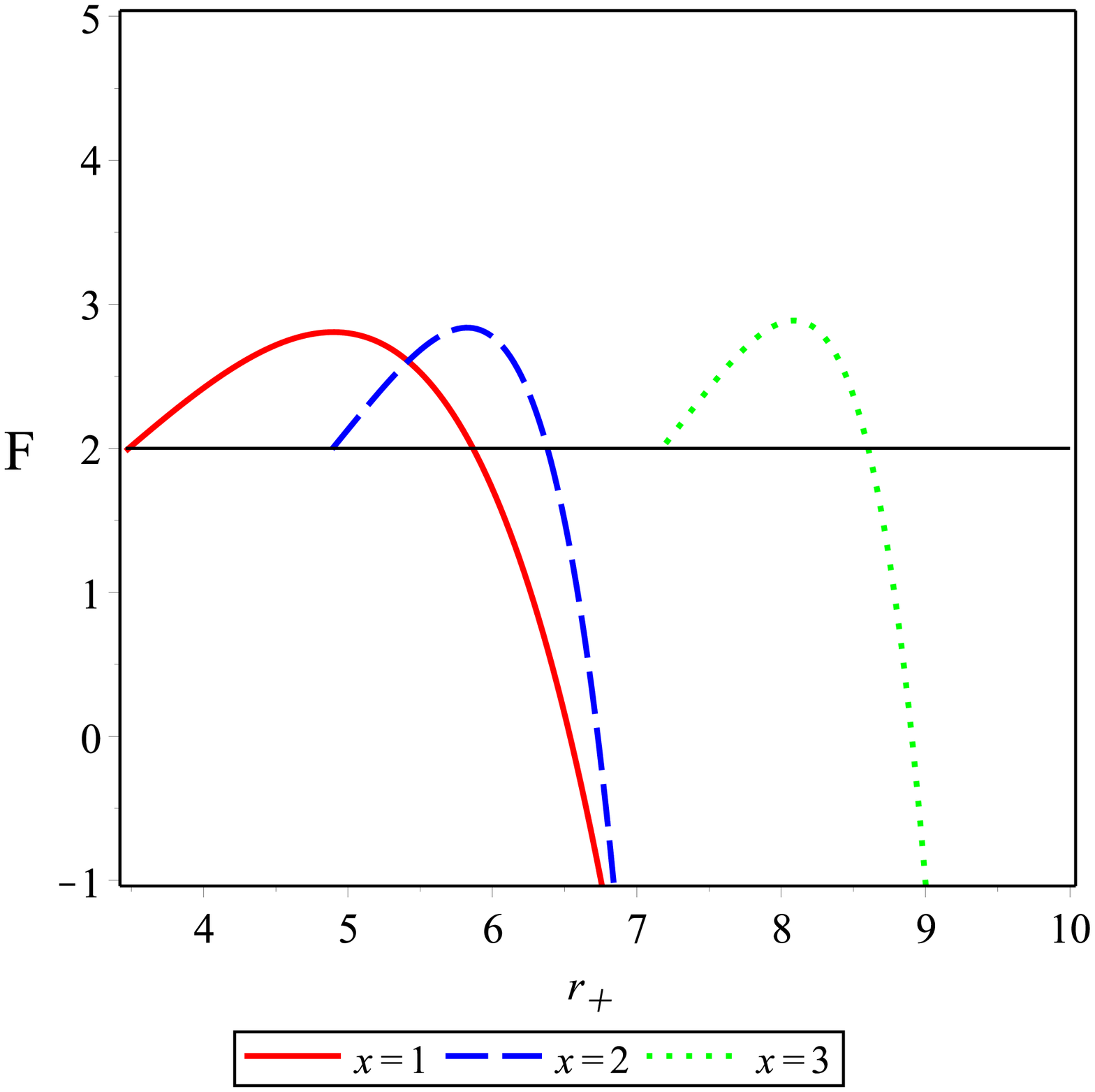}\label{fig52}}\hspace*{.5cm}
\subfigure[ $F-r_+$(red), $0.11F-r_+$(blue) and $0.008F-r_+$(green) for $\alpha=1$, $x=1$ and $G=1$.
    ]{\includegraphics[scale=0.25]{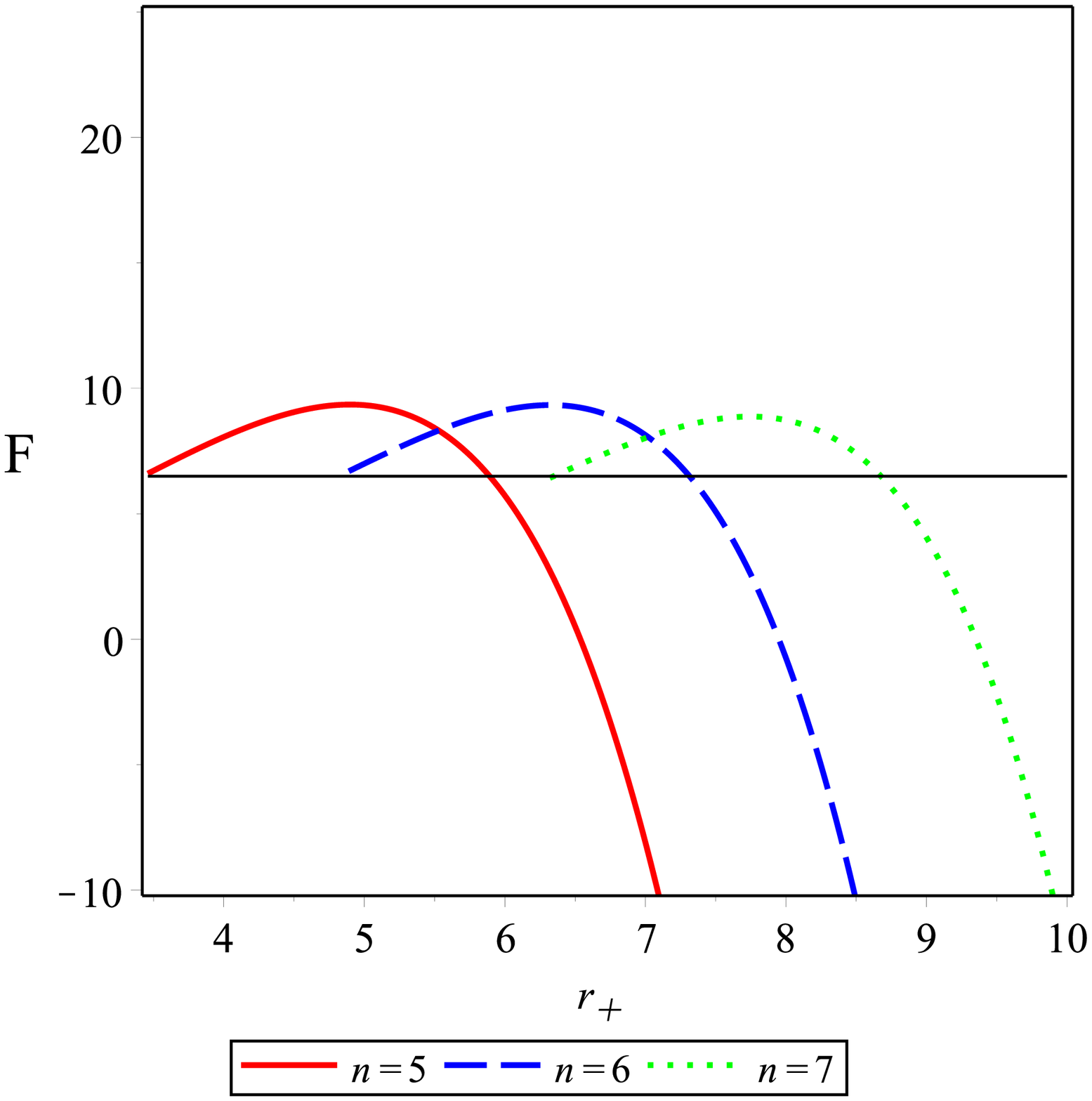}\label{fig53}}
\caption{Behavior of the free energy versus $r_+$ for $k=-1 $ for "$-$" sign
branch.}
\label{fig5}
\end{figure}

\begin{figure}[h!]
\subfigure[ $7\times10^{-3}F-T$(red), $10^{-3}F-T$(blue) and
$2\times10^{-5}F-T$(green) for $n=7$, $x=1$ and
$G=1$.]{\includegraphics[scale=0.25]{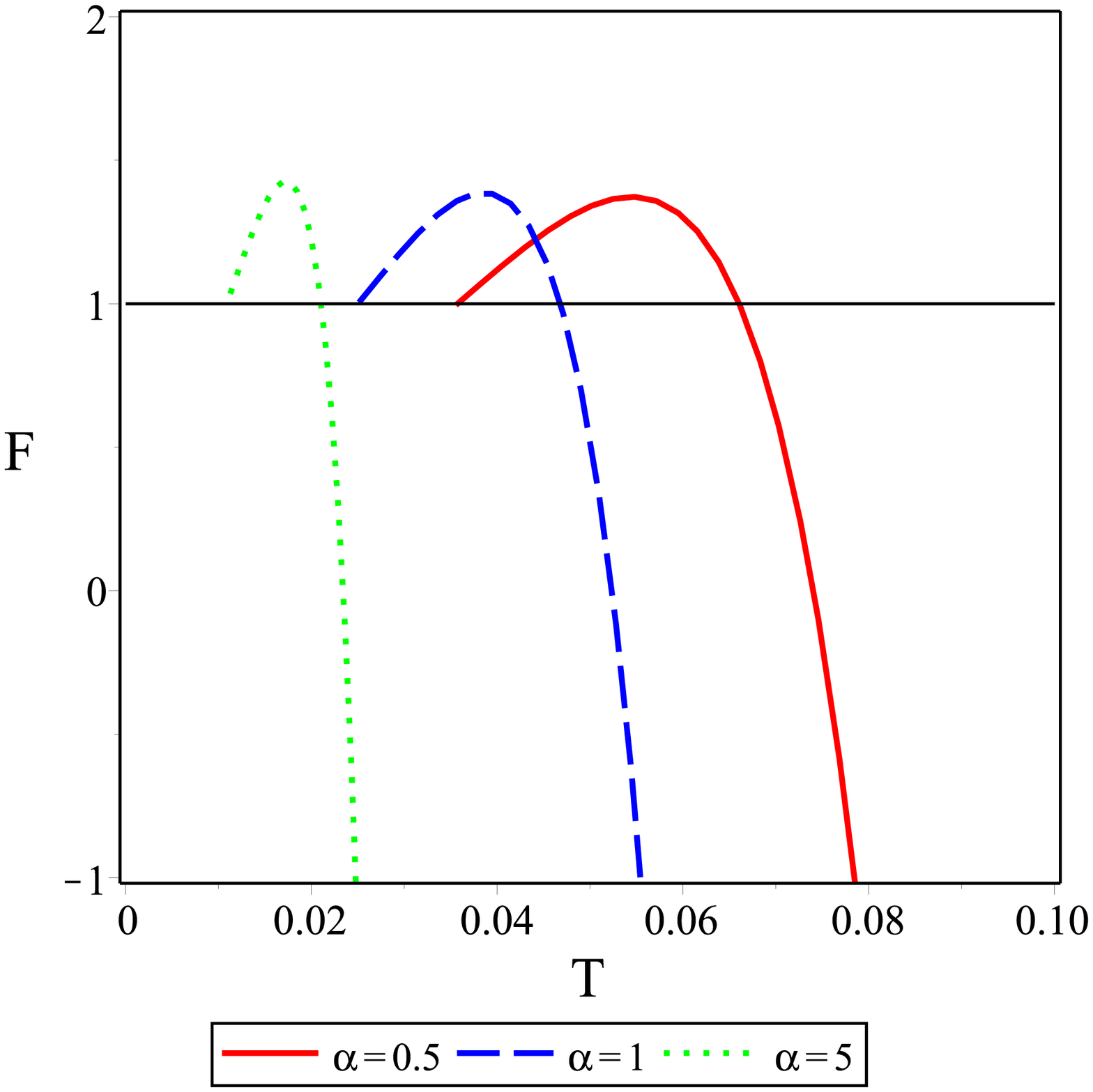}\label{fig61}} \hspace*{.5cm}
\subfigure[ $F-T$(red), $3\times10^{-5}F-T$(blue) and $2\times10^{-11}F-T$(green) for $n=5$, $\alpha=1$ and $G=1$.
    ]{\includegraphics[scale=0.25]{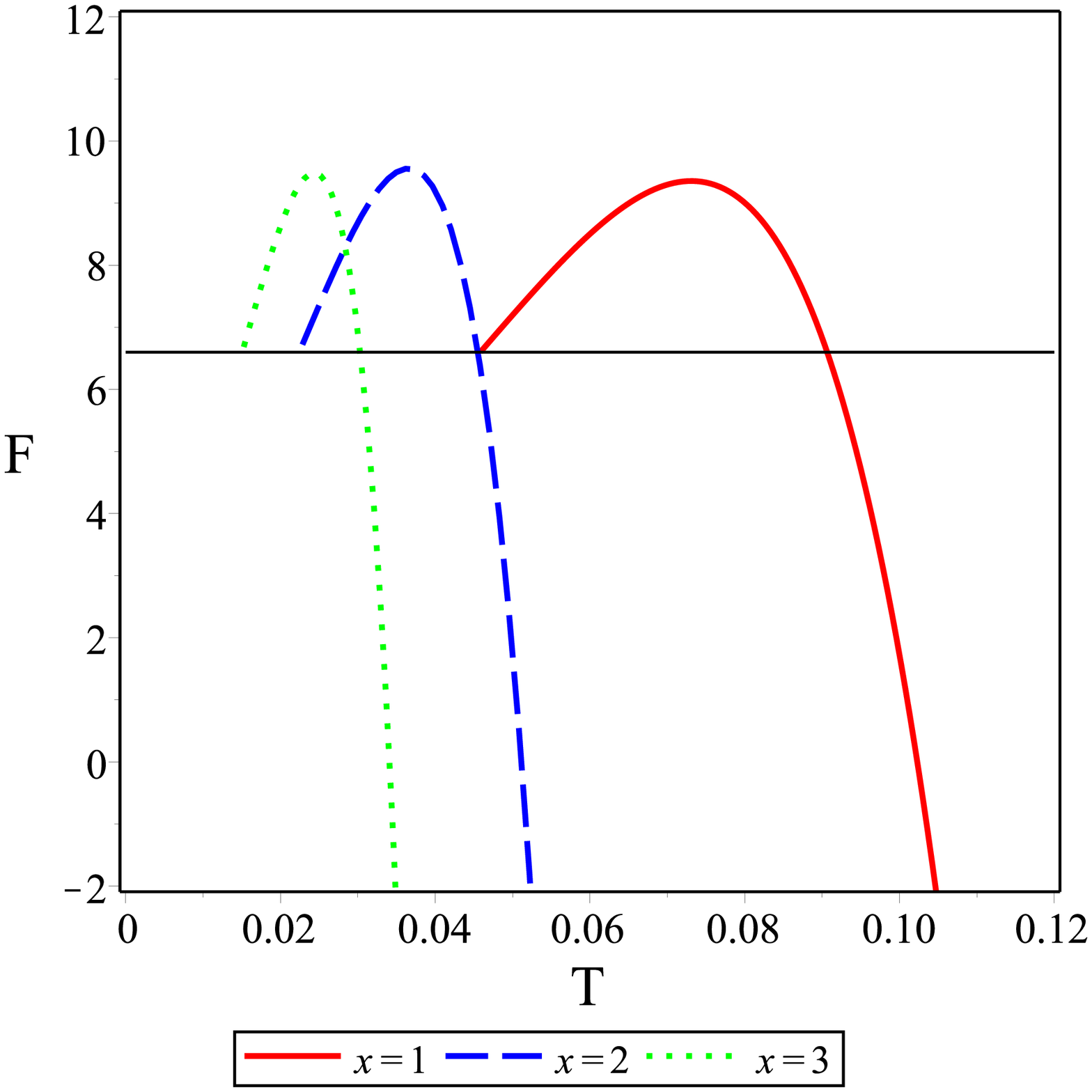}\label{fig62}}\hspace*{.5cm}
\subfigure[ $F-T$(red), $0.11F-T$(blue) and $0.008F-T$(green) for $\alpha=1$, $x=1$ and $G=1$.
    ]{\includegraphics[scale=0.25]{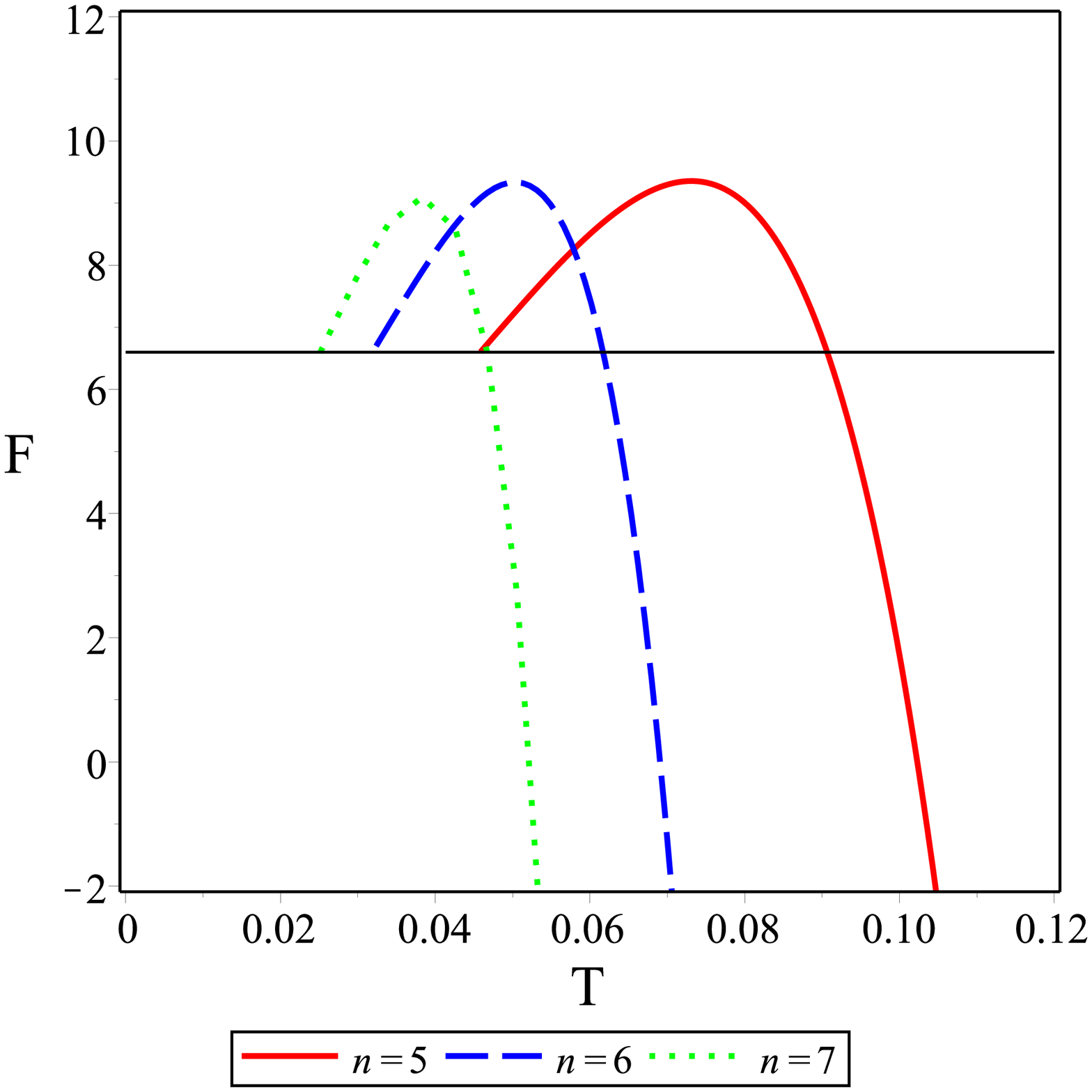}\label{fig63}}
\caption{Behavior of the free energy versus $T$ for $k=-1 $ for "$-$" sign
branch.}
\label{fig6}
\end{figure}

In order to investigate how the free energy behaves by changing the
parameters of the model, we have provided Fig. \ref{fig5} and Fig. \ref{fig6}%
. Figure \ref{fig5} indicates that by increasing the value of parameter $%
\alpha$ \ref{fig51}, parameter $x$ \ref{fig52} and dimension of spacetime $n$
\ref{fig53} (and fixed values of the other parameters), the maximum value of
this function as well as the radius in which this maximum value occurs
increases. In addition, the behavior of the free energy versus $T$ is
depicted in Fig. \ref{fig6}. As this figure shows, increasing each of the
parameters $\alpha$, $x$ and $n$ causes the maximum value of the free energy
function to occur at a lower temperature.

\vspace{0.5cm} ${\bullet{\mathbf{{\mathbf{Case\; k=1}: }}}} $ \vspace{0.2cm}

Regarding the solution with $k=1$, the temperature and the heat capacity
have the following form
\begin{equation}
T={\frac {nxr_{+}^{x}}{8\pi\,\sqrt { \tilde \alpha ^{x+1}}} \left( 1+{\frac {
( n-4) \tilde \alpha ^{x}}{n{x}^{2}\,r_{+}^{2x}}} \right) },
\end{equation}
\begin{equation}
C={\frac {{x}^{2}( n-1) r_{+}^{( n-1 ) x}\sqrt { \tilde \alpha ^{x-1}}}{G}
\left( 1+{\frac {( n-4) \tilde \alpha ^{x}}{r_{+} ^{2x}n{x}^{2}}} \right)
\left( 1+{\frac { \tilde \alpha ^{x}}{{x}^{2}r_{+}^{2x} }} \right) \left( 1-{%
\frac {( n-4) \tilde\alpha ^{x}}{r_{+}^{2x }n{x}^{2}}} \right) ^{-1}}.
\end{equation}

As it is clear from the above relations, although the temperature is always
positive for each value of model parameters, the heat capacity meets a
positive value just for a range of horizon radii. Moreover, the heat
capacity function experiences a divergence at
\begin{figure}[h!]
\centering\includegraphics[scale=0.4]{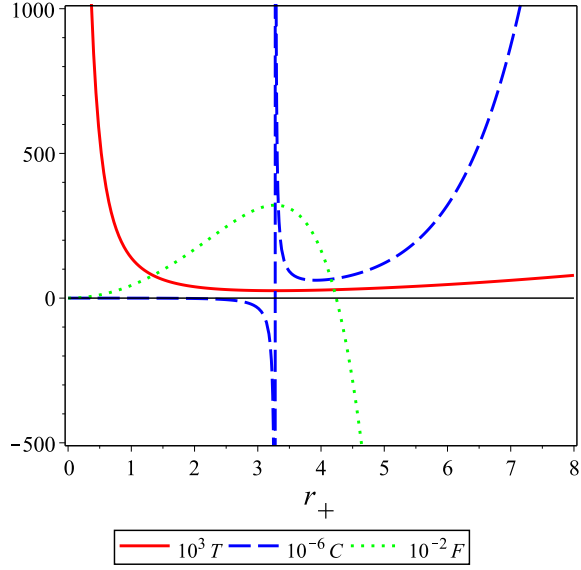}
\caption{Behavior of $C$, $T$ and $F$ with respect to $r_+$ for $k=1$, $n=5$%
, $x=2 $ and $\protect\alpha=1$ for "$-$" sign branch.}
\label{fig7}
\end{figure}
\begin{equation}
r_{{\tiny {\text{div}}}}=\left( {\frac {( n-4) \tilde \alpha ^{x}}{n{x}^{2}}}
\right) ^{\frac{1}{2\,x}},
\end{equation}
which its position alters by changing the value of the model parameters. To
be more clear, we have provided Fig. \ref{fig7} in which the behavior of the
temperature, heat capacity and free energy are illustrated for a set of
model parameters. Hence, if the radius of the event horizon is larger than $%
r_{{\tiny {\text{div}}}}$, the black hole is thermally stable and enjoys
some necessary criteria for viable solutions. In addition, Fig. \ref{fig77}
states that increasing the value of parameters $n$, $x$ and $\alpha$ leads
to increasing the amount of $r_{{\tiny {\text{div}}}}$ and causes the
stability interval of the black hole to start from a larger radius.

\begin{figure}[h!]
\subfigure[ $C-r_+$(red), $10^{-2}C-r_+$(blue) and $10^{-4}C-r_+$(green) for
$\alpha=1$, $x=1$ and
$G=1$.]{\includegraphics[scale=0.25]{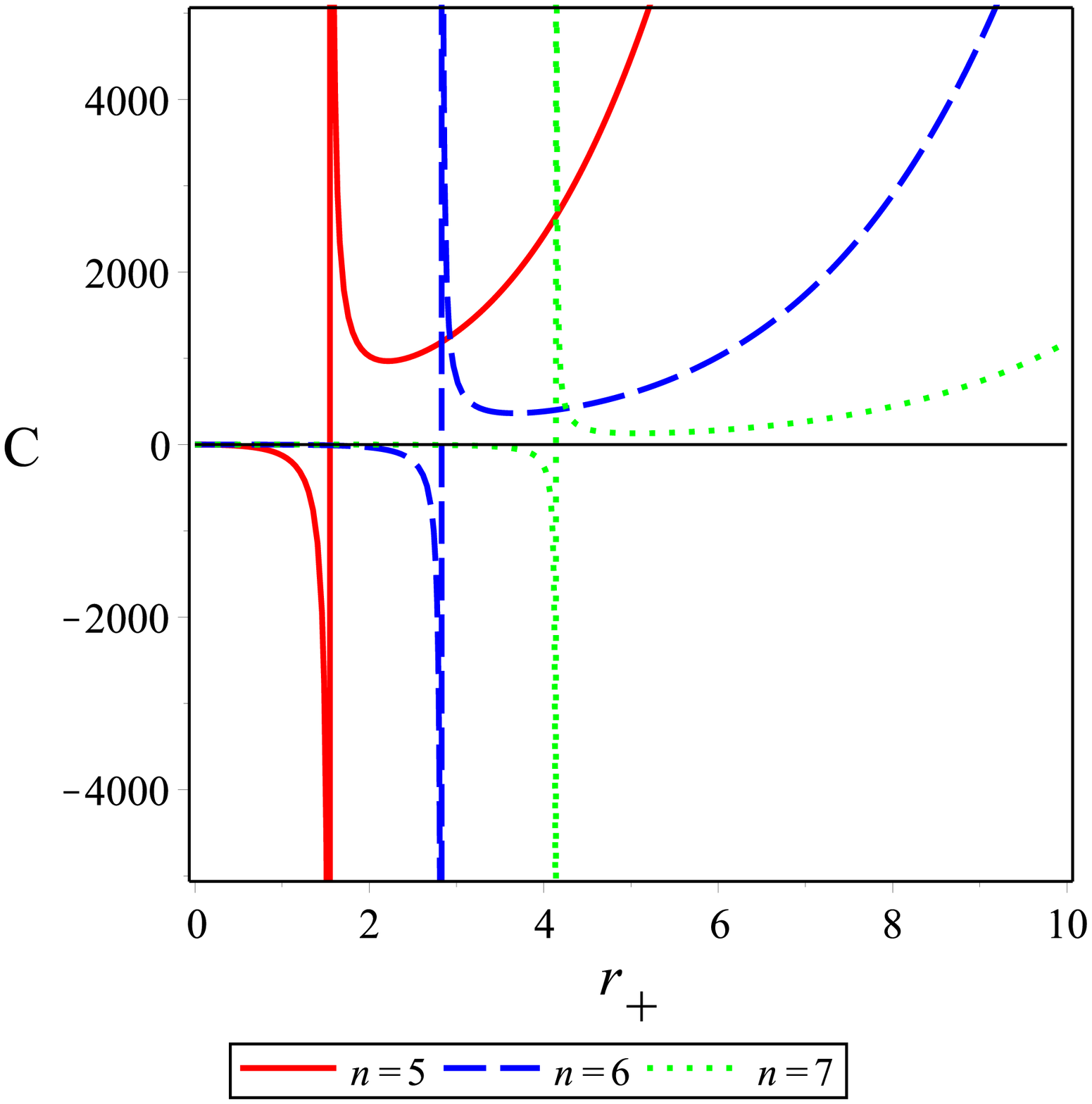}\label{fig71}} \hspace*{.5cm}
\subfigure[ $10C-r_+$(red), $10^{-4}C-r_+$(blue) and $10^{-10}C-r_+$(green) for $n=5$, $\alpha=1$ and $G=1$.
    ]{\includegraphics[scale=0.25]{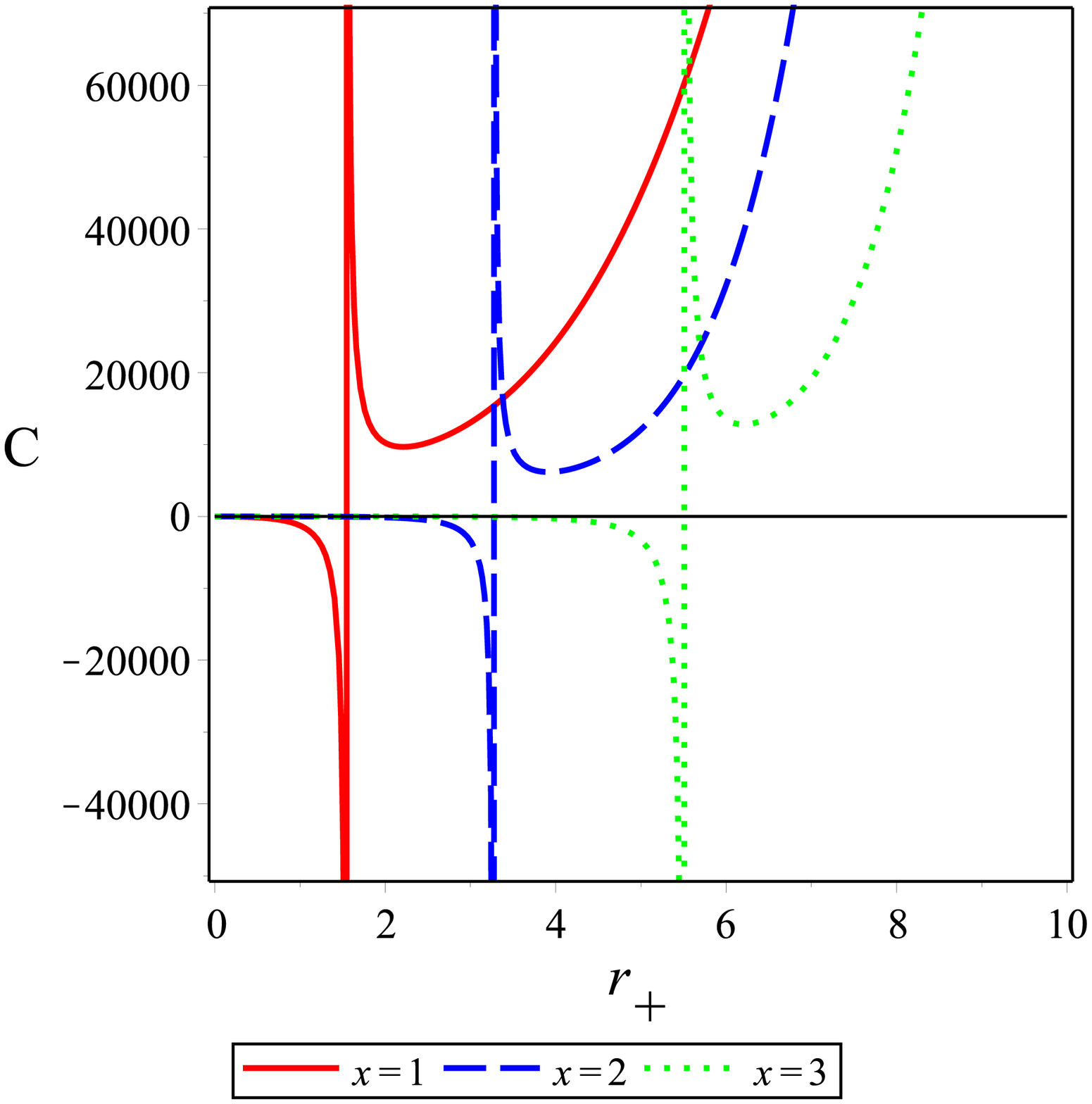}\label{fig72}}\hspace*{.5cm}
\subfigure[ $10\,C-r_+$(red), $C-r_+$(blue) and $10^{-2}\,C-r_+$(green) for $n=5$, $x=2$ and $G=1$.
    ]{\includegraphics[scale=0.25]{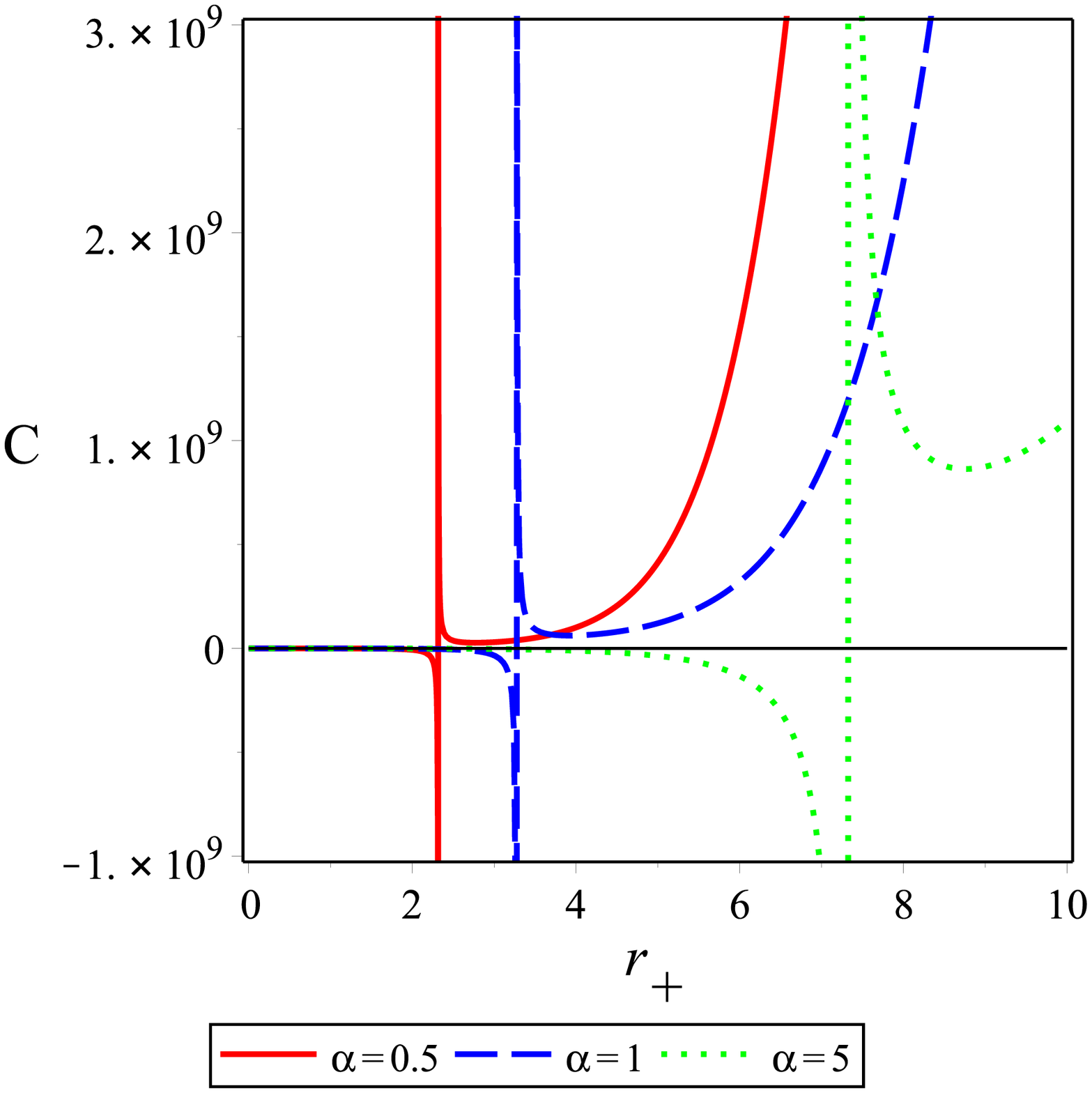}\label{fig73}}
\caption{Behavior of the heat capacity versus $r_+$ for $k=1 $ for "$-$"
sign branch.}
\label{fig77}
\end{figure}

\begin{figure}[h!]
\subfigure[ $F-r_+$(red), $F-r_+$(blue) and $0.1F-r_+$(green) for
$\alpha=1$, $x=1$ and
$G=1$.]{\includegraphics[scale=0.25]{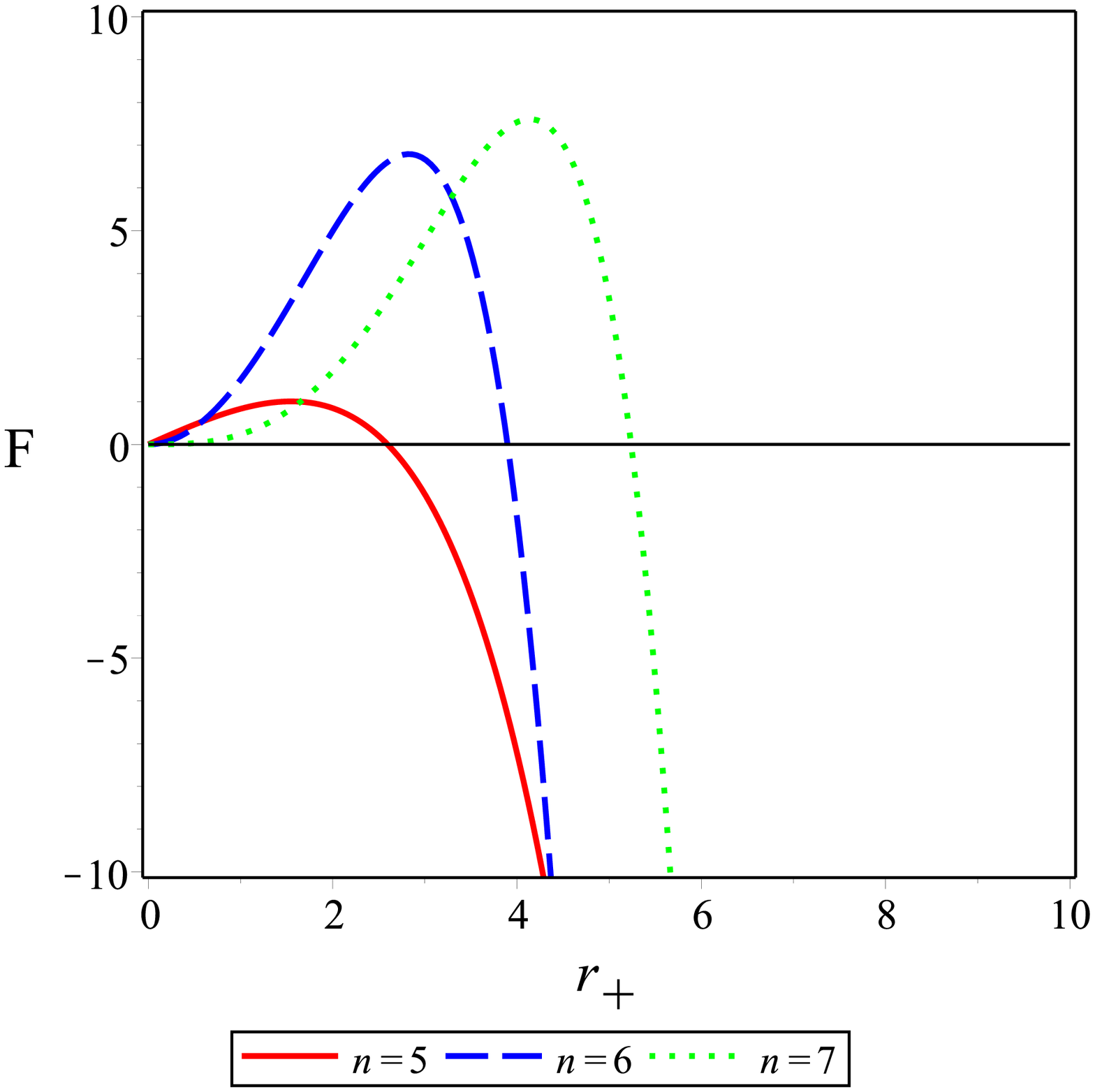}\label{fig81}} \hspace*{.5cm}
\subfigure[ $F-r_+$(red), $10^{-4}F-r_+$(blue) and $10^{-10}F-r_+$(green) for $n=5$, $\alpha=1$ and $G=1$.
    ]{\includegraphics[scale=0.25]{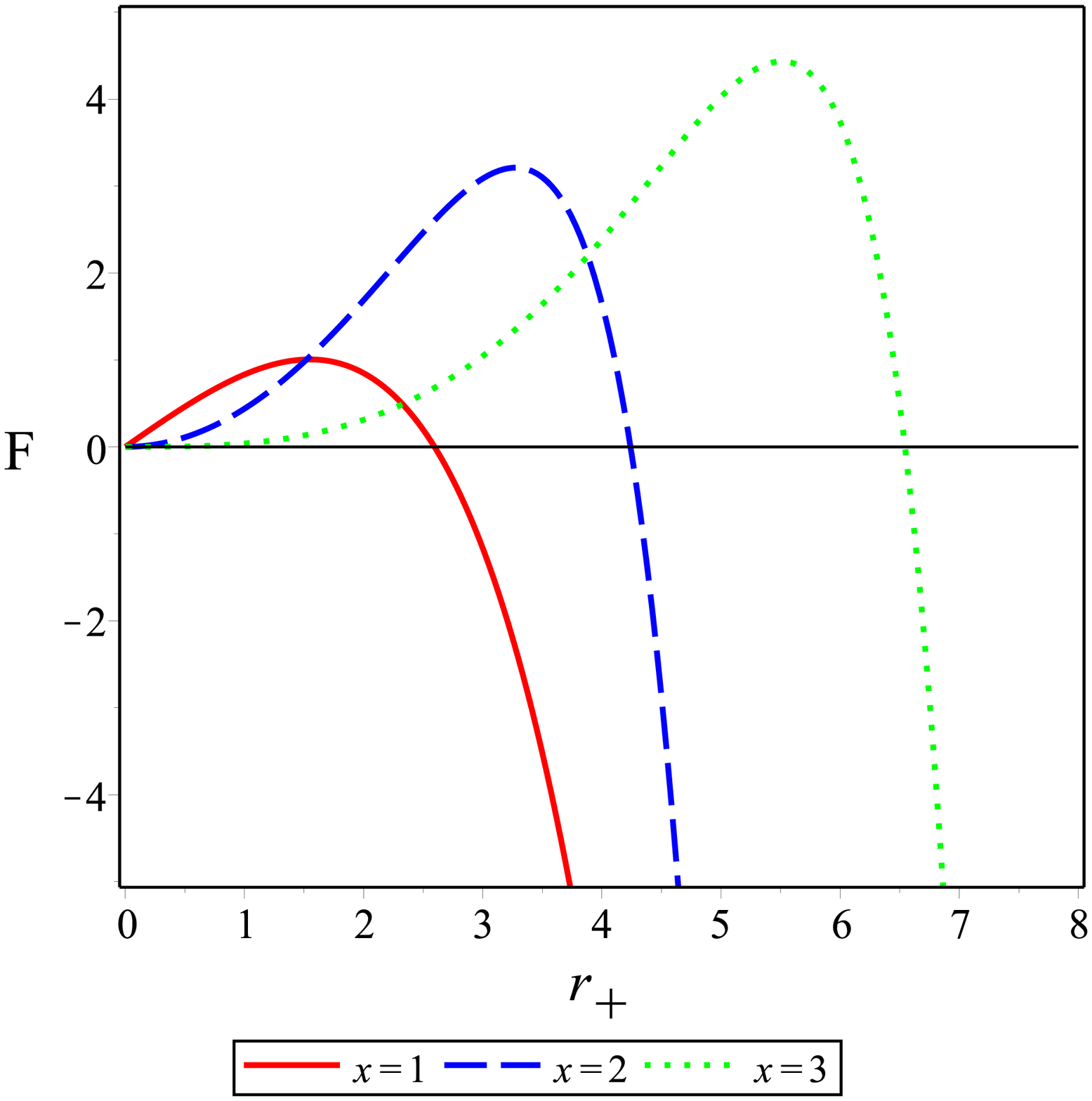}\label{fig82}}\hspace*{.5cm}
\subfigure[ $10^{-3}F-r_+$(red), $10^{-4}F-r_+$(blue) and $2\times10^{-7}F-r_+$(green) for $n=5$, $x=2$ and $G=1$.
    ]{\includegraphics[scale=0.25]{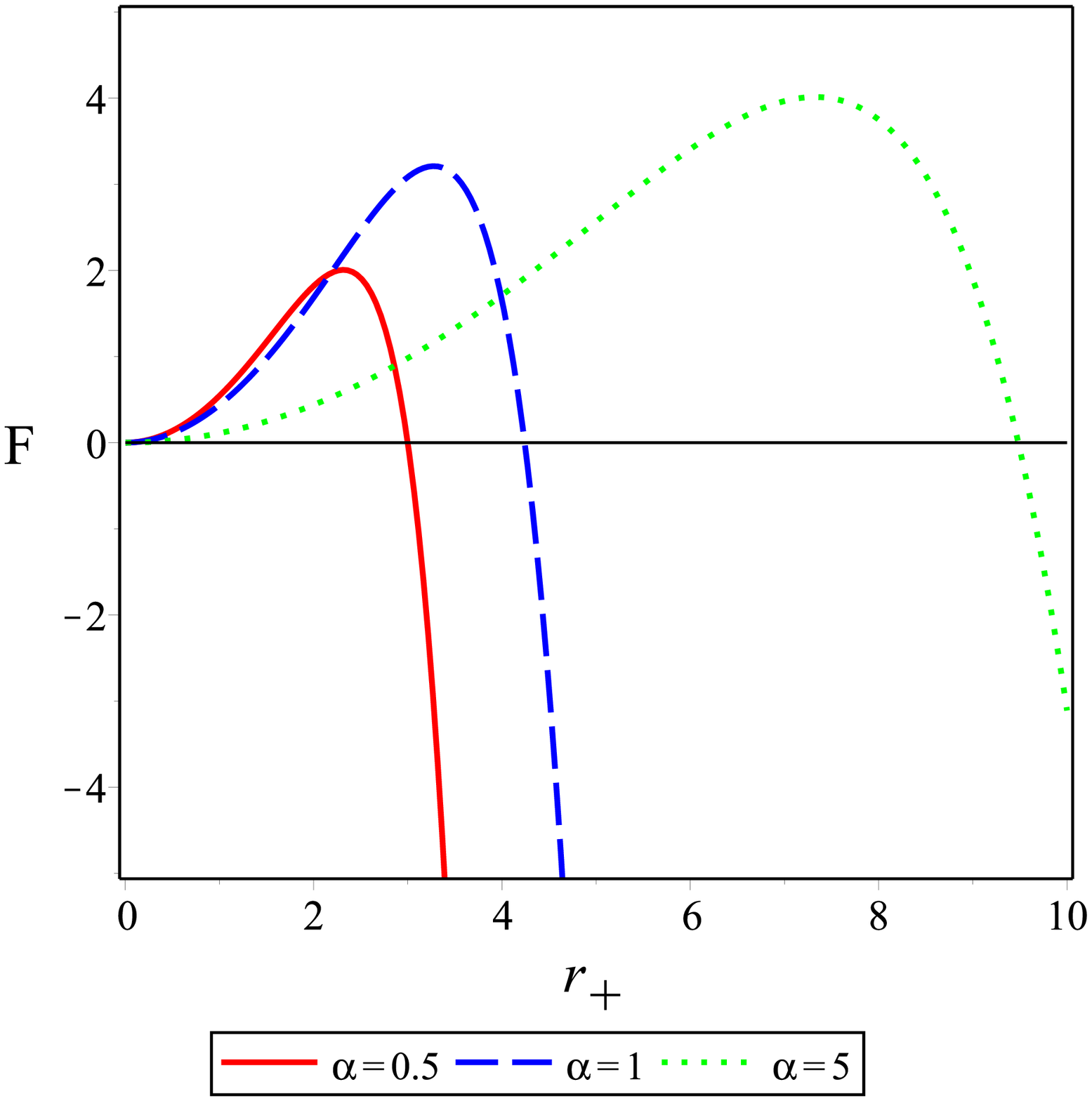}\label{fig83}}
\caption{Behavior of the free energy versus $r_+$ for $k=1$ for "$-$" sign
branch.}
\label{fig8}
\end{figure}

Similar to the case of $k=-1$, the energy function always meets a positive
maximum value at $r_+=r_{{\tiny {\text{div}}}}=\left( {\frac { \tilde
\alpha^{x} \left( n-4 \right) }{n{x}^{2}}} \right) ^{\frac{1}{2\,x}} $ and
tends to negative infinity when $r_+ \rightarrow\,\infty$ (see Fig. \ref%
{fig7}). Therefore, the maximum value of the free energy occurs at a horizon
radius in which the heat capacity diverges. Furthermore, the maximum value
of this function as well as the radius in which this maximum value occurs
increases by increasing the value of parameters $x$, $n$ and $\alpha$ (see
Fig. \ref{fig8} for more details).

\section{Positive sign branch}

\label{plus}

Considering the branch of the metric function \eqref{Msol} with "+" sign, we
can easily find that the metric functions associated with $k=-1$ have real
positive non-zero roots and hence black holes could only exist for the
pseudo-hyperbolic horizon (other curvatures result in a naked singularity
and do not have black hole interpretation).

\begin{figure}[tbp]
\centering
\subfigure[x=2]{\includegraphics[scale=0.3]{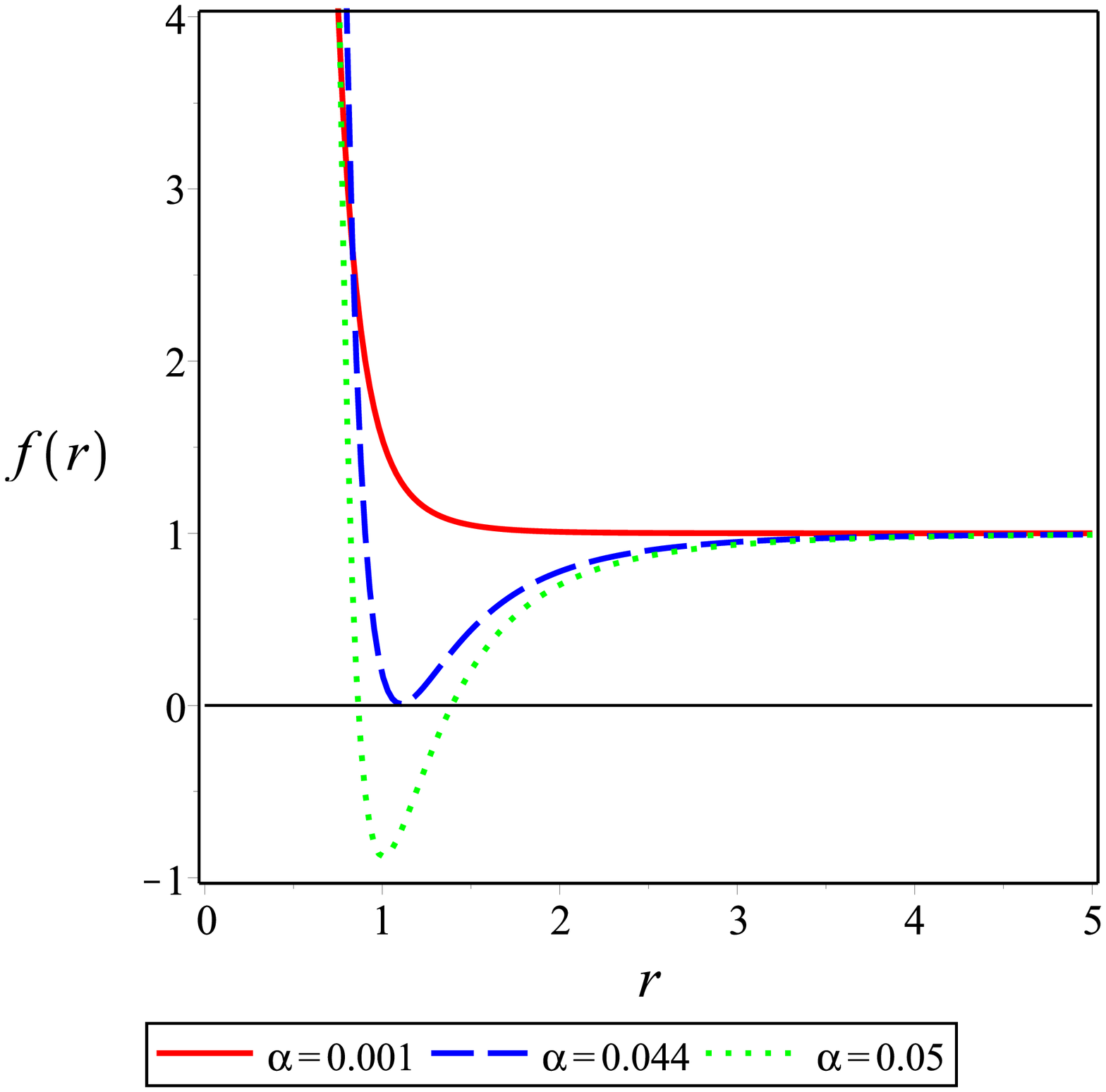}\label{fig3-1}} \hspace*{%
.1cm}
\subfigure[$\alpha=0.044$
    ]{\includegraphics[scale=0.3]{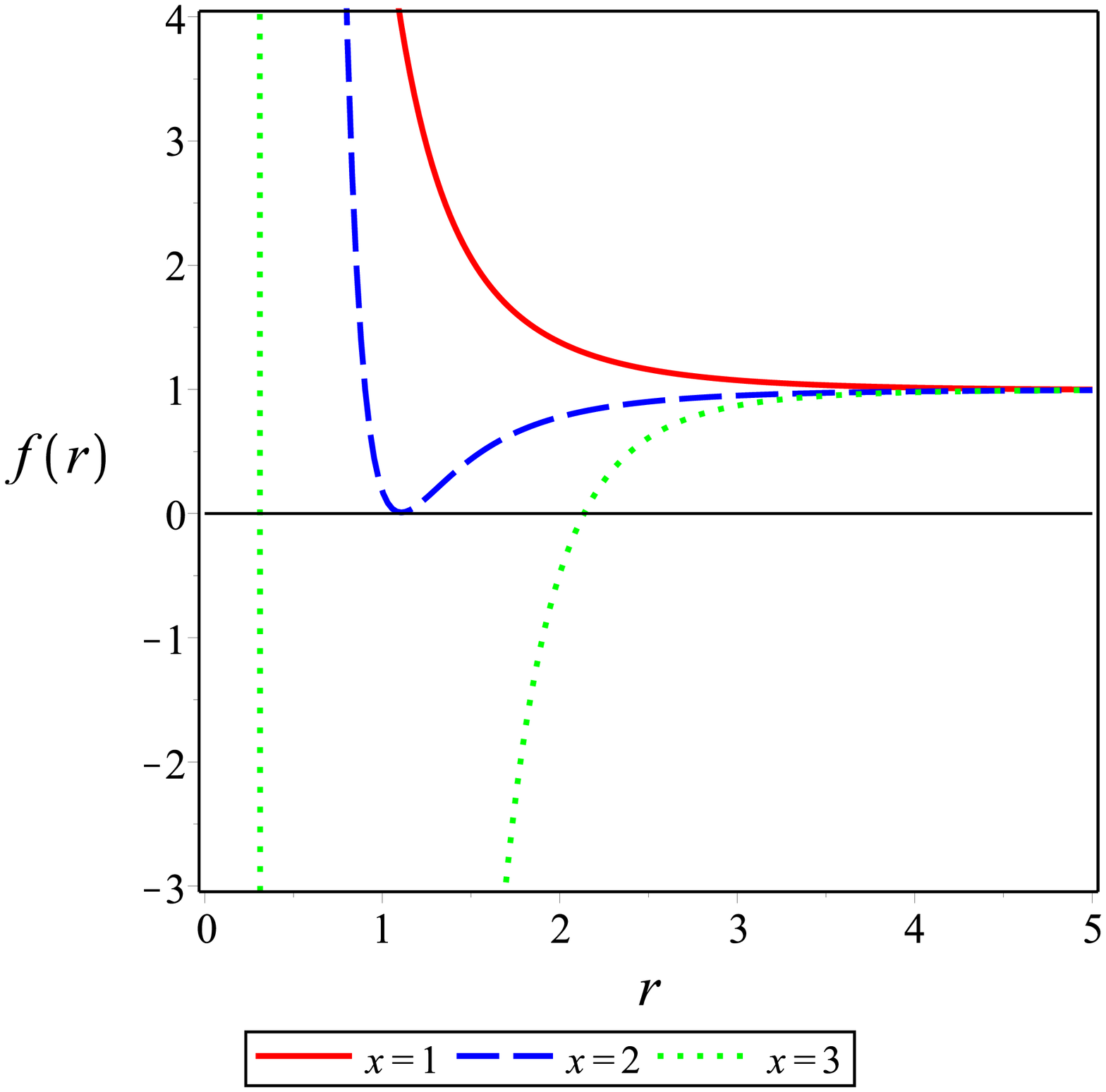}\label{fig3-2}}
\caption{Behavior of $f(r)$ versus $r$ for $k=-1$, $n=6$, $G=1$ and $M=5$
for "$+$" sign branch.}
\label{fig3}
\end{figure}

Now, we try to investigate physical properties of the solution with the
pseudo-hyperbolic horizon by studying the behavior of the obtained solution
and looking for the horizons. To this end, we have plotted the function $%
f(r) $ versus $r$ for different model parameters in Fig. \ref{fig3}. These
figures show that, depending on the metric parameters, this solution could
represent a black hole with two horizons, an extreme black hole with a
degenerate horizon, or a naked singularity. Figure \ref{fig3-1} indicates
that by increasing the value of the Gauss-Bonnet coefficient, $\alpha$, (and
fixed values of the other parameters), the number of horizons changes from
zero to two. In Fig. \ref{fig3-2} we consider the effect of the dynamical
exponent $x$ on the number of the horizons and find that increasing the value
of the dynamical exponent $x$ leads to increasing the number of horizons to
two horizons.

To ensure the number of horizons, one can calculate $f(r)$ derivatives'
roots. As mentioned earlier, since the number of $f(r)$ derivatives' roots
shows the number of function's extrema, the existence of more than one
extrema for $f(r)$ indicates that the number of horizons could be more than
two. However, the existence of only one extremum represents that the maximum
number of horizons will be two. Concerning our solution, we find that
\begin{equation}
r|_{f^{\prime }(r)=0}=\left({\frac {1}{\tilde{\alpha}^2}\sqrt[x]{{\frac {{n}%
^{2}\tilde{\alpha}MG\pi{x} }{2(n-1)}}}} \right) ^{ \frac{1}{n-4}}.
\end{equation}
Since there is only one acceptable extremum (real and positive), independent
of the metric parameters $x$, $n$, and $\alpha$, it can be ensured that the
maximum number of horizons is two.

Concerning the black hole singularity, our calculations show that the
Kretschmann invariant and Ricci scalar follow similar relations to the Eqs. (%
\ref{Kres}) and (\ref{Ricci}). Therefore, this solution also has an
essential singularity at the origin. Besides, considering Fig. \ref{fig3},
it is clear that the singularity is timelike and the behavior of the
solutions is similar to charged black holes.

\subsection{Thermodynamic properties}

\label{plusA}

Now, we study the thermodynamic structure of the solution with "$+$" sign.
Similar to the previous section, we first obtain the thermodynamic
quantities and then investigate the local and global stability of the system
by calculating heat capacity and free energy. As was already mentioned, for
the solution with "$+$" sign, black holes can be found only for the
pseudo-hyperbolic horizon. So, we conduct our investigation for the case of $%
k=-1$.

The mass of the black hole is obtained as
\begin{equation}  \label{mass:2}
M=\frac{(n-1)r_{+}^{(n-2)x}}{8\pi Gx}\frac{\tilde\alpha^{x}}{\hat{\alpha}}%
\left[ -1+\frac{\tilde\alpha^{x}}{2x^{2}r_{+}^{2x}}+\frac{x^{2}r_{+}^{2x}}{%
2\tilde\alpha^{x}}\right] .
\end{equation}

From Eq. (\ref{temp}), the temperature is given by

\begin{eqnarray}
T_{{\tiny {H}}}={\frac {4\,{x}^{2}r_{+}^{2x}- \left| {x}^{2}r_{+} ^{2\,x}+k
\tilde{\alpha} ^{x} \right|( n-4) }{8\pi\,x\,r_{+}^{x}\sqrt { \tilde{\alpha}
^{x+1}}}}.  \label{Temp22new}
\end{eqnarray}

Considering the condition $r_{+}>\left( {\frac { \tilde{\alpha} ^{x}}{{x}^{2}%
}} \right) ^{\frac{1}{2x}}$, one finds
\begin{equation}
T=\frac{(n-8)xr_{+}^{x}}{8\pi \sqrt{\tilde\alpha^{x+1}}}\left(-1+ \frac{%
(n-4)\tilde\alpha^{x}}{(n-8)x^{2}r_{+}^{2x}}\right) .  \label{Temp:2}
\end{equation}

It is worth pointing out that for $5\leq n \leq8$, the temperature
is always positive. However, for $n\geq 9$, it is positive only in
special region,
$\left( {\frac { \tilde{\alpha} ^{x}}{{x}^{2}}} \right) ^{\frac{1}{2x}%
}<r_{+}<\left( {\frac { (n-4)\tilde{\alpha} ^{x}}{(n-8){x}^{2}}} \right) ^{%
\frac{1}{2x}}$, meaning that in higher dimensions, we have a
restriction for the horizon radius of the physical black holes in
the theory under consideration.

\begin{figure}[h!]
\subfigure[ $10^{-3}C-r_+$ (red), $10^{-8}C-r_+$ (blue) and
$10^{-15}C-r_+$ (green) for $n=6$ and
$\alpha=0.5$.]{\includegraphics[scale=0.25]{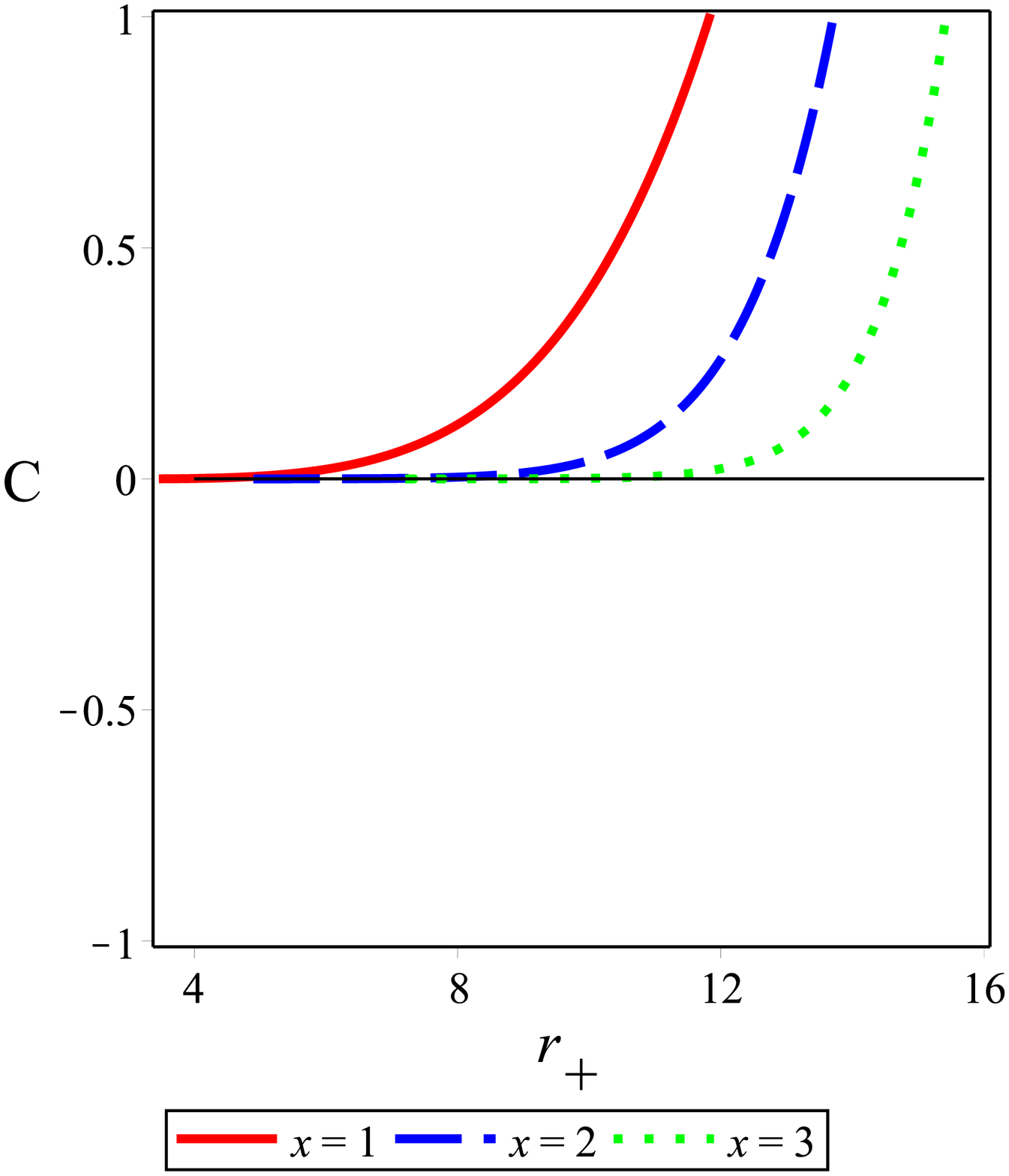}\label{C61}} \hspace*{.5cm}
\subfigure[ $10^{-4}C-r_+$ (red), $10^{-10}C-r_+$
(blue) and $10^{-20}C-r_+$ (green) for $n=7$ and $\alpha=0.5$.
    ]{\includegraphics[scale=0.25]{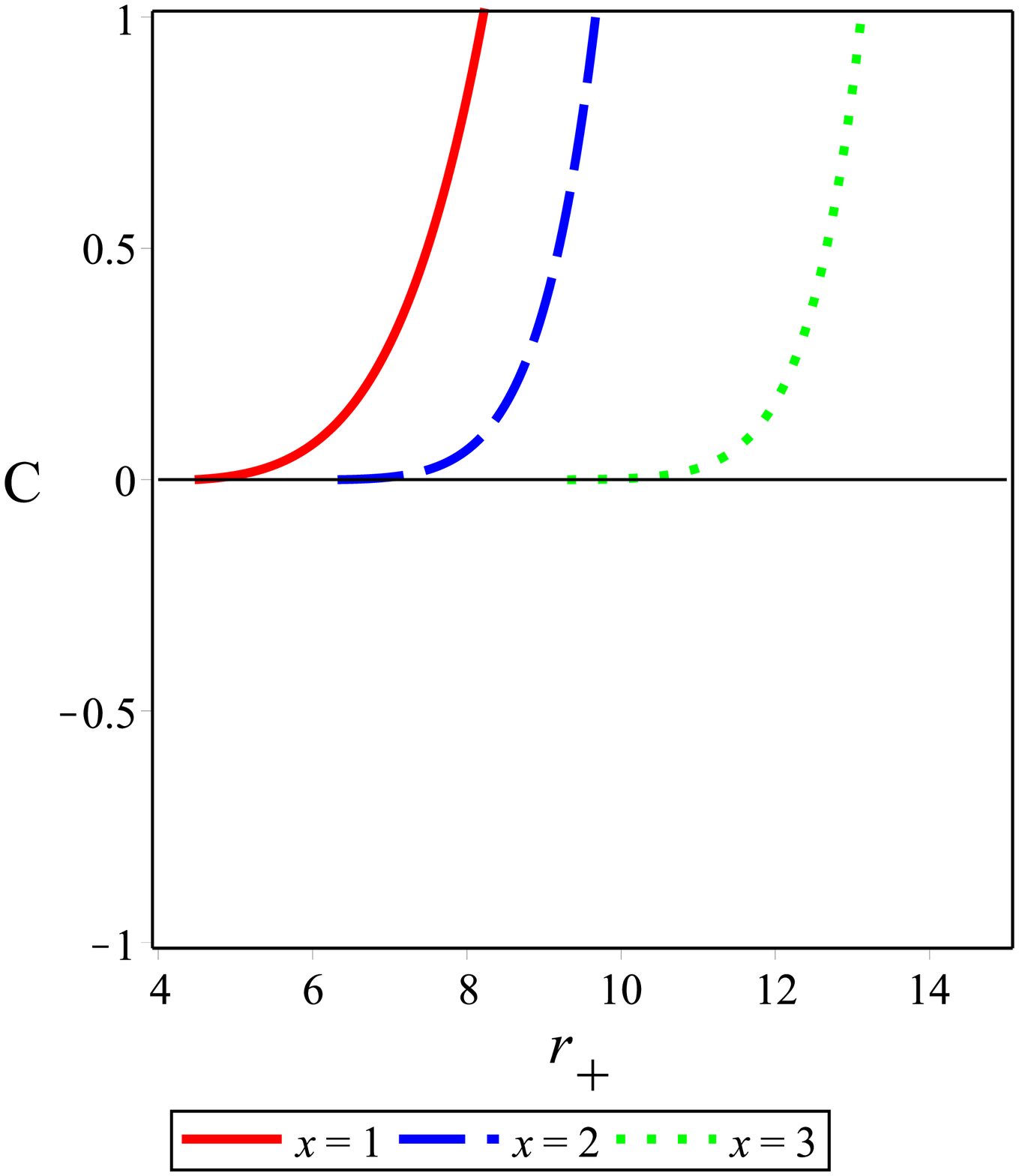}\label{C71}}\hspace*{.5cm}
\subfigure[ $10^{-5}C-r_+$(red), $10^{-15}C-r_+$(blue) and
$10^{-25}C-r_+$(green) for $n=8$ and $\alpha=0.5$.
    ]{\includegraphics[scale=0.25]{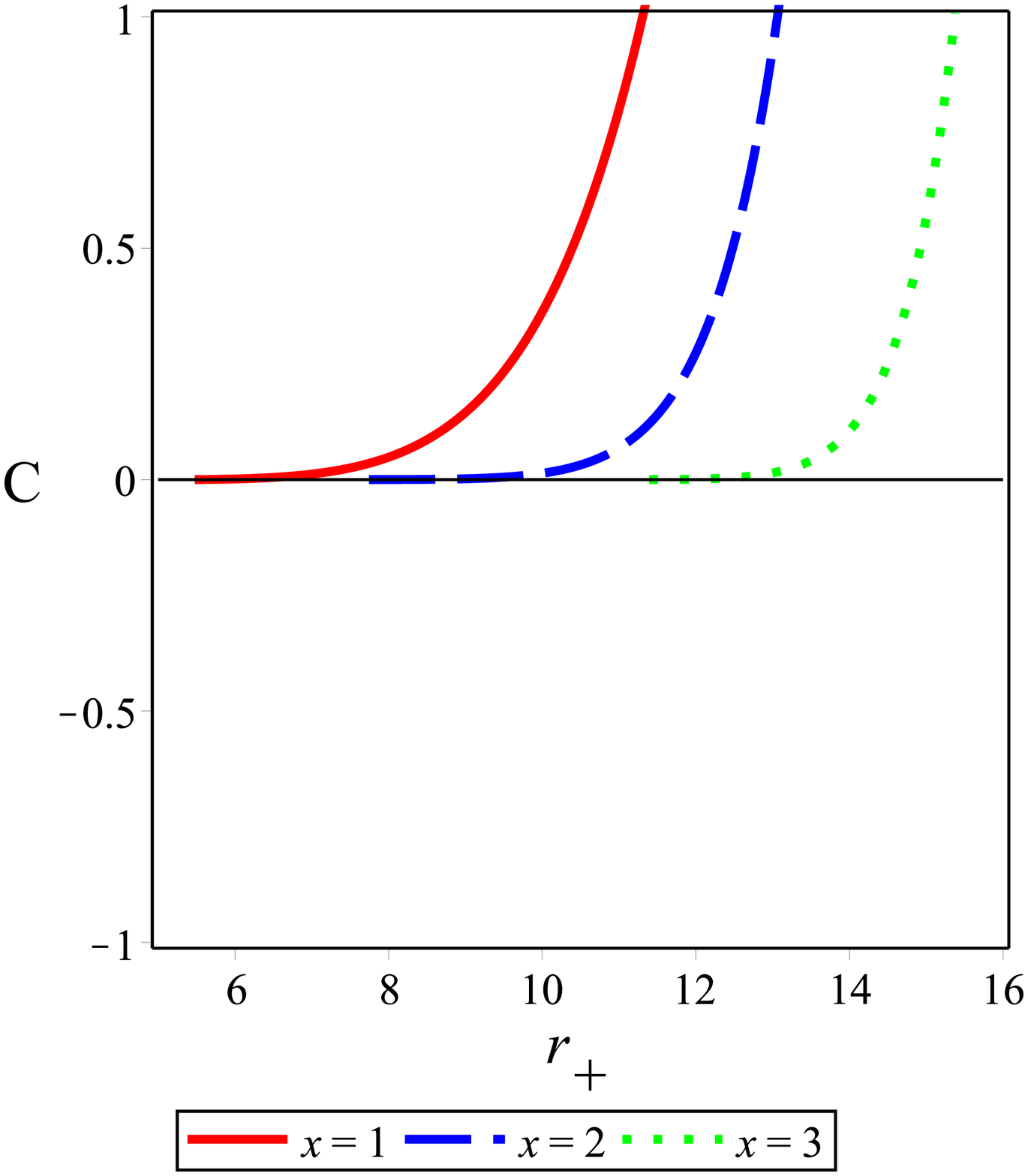}\label{C81}}\newline
\subfigure[ $10^{-8}C-r_+$ (red), $10^{-10}C-r_+$ (blue) and
$10^{-12}C-r_+$ (green) for $n=6$ and
$x=2$.]{\includegraphics[scale=0.25]{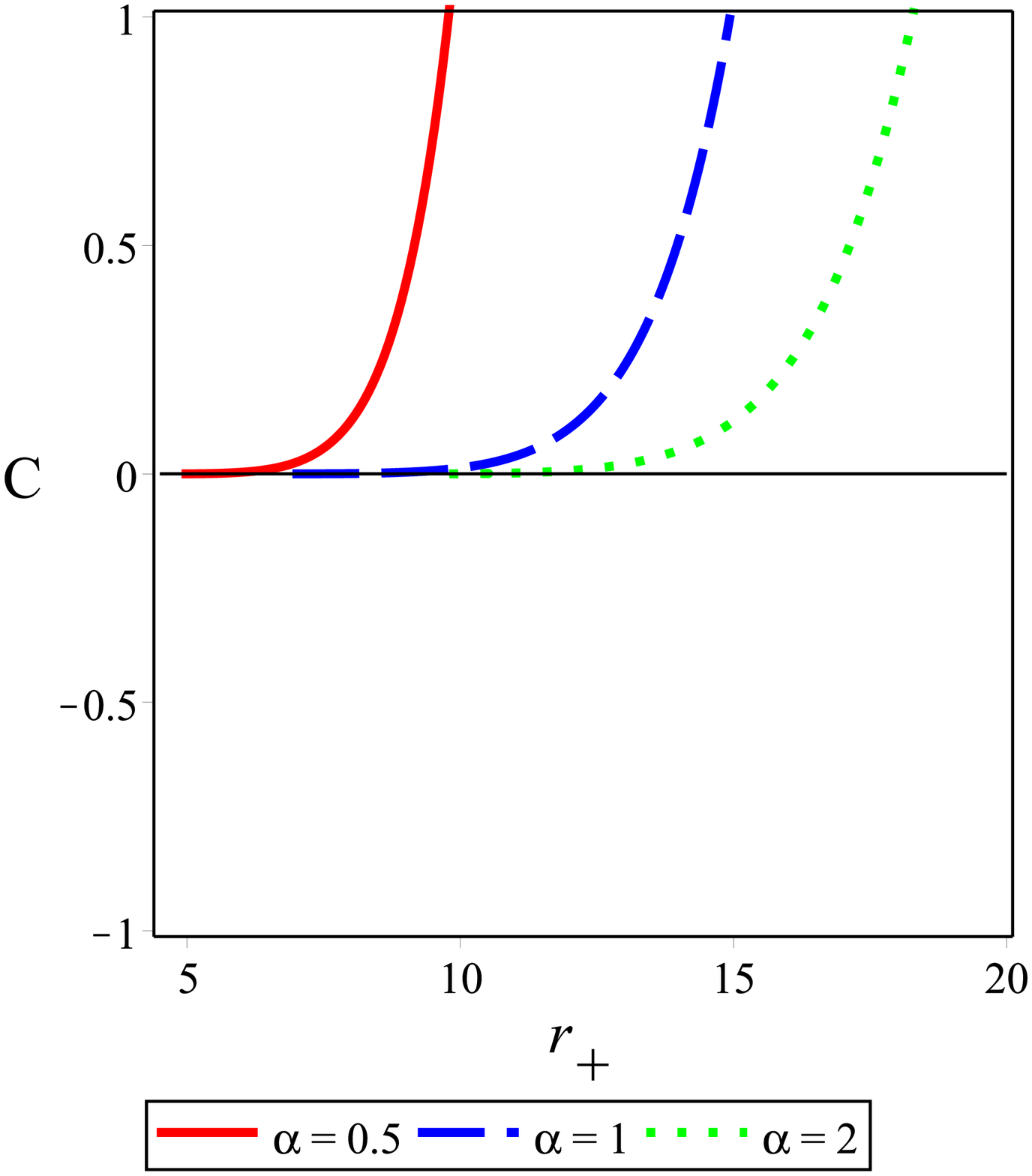}\label{C62}} \hspace*{.5cm}
\subfigure[ $10^{-11}C-r_+$ (red), $10^{-13}C-r_+$
(blue) and $10^{-16}C-r_+$ (green) for $n=7$ and $x=2$.
    ]{\includegraphics[scale=0.25]{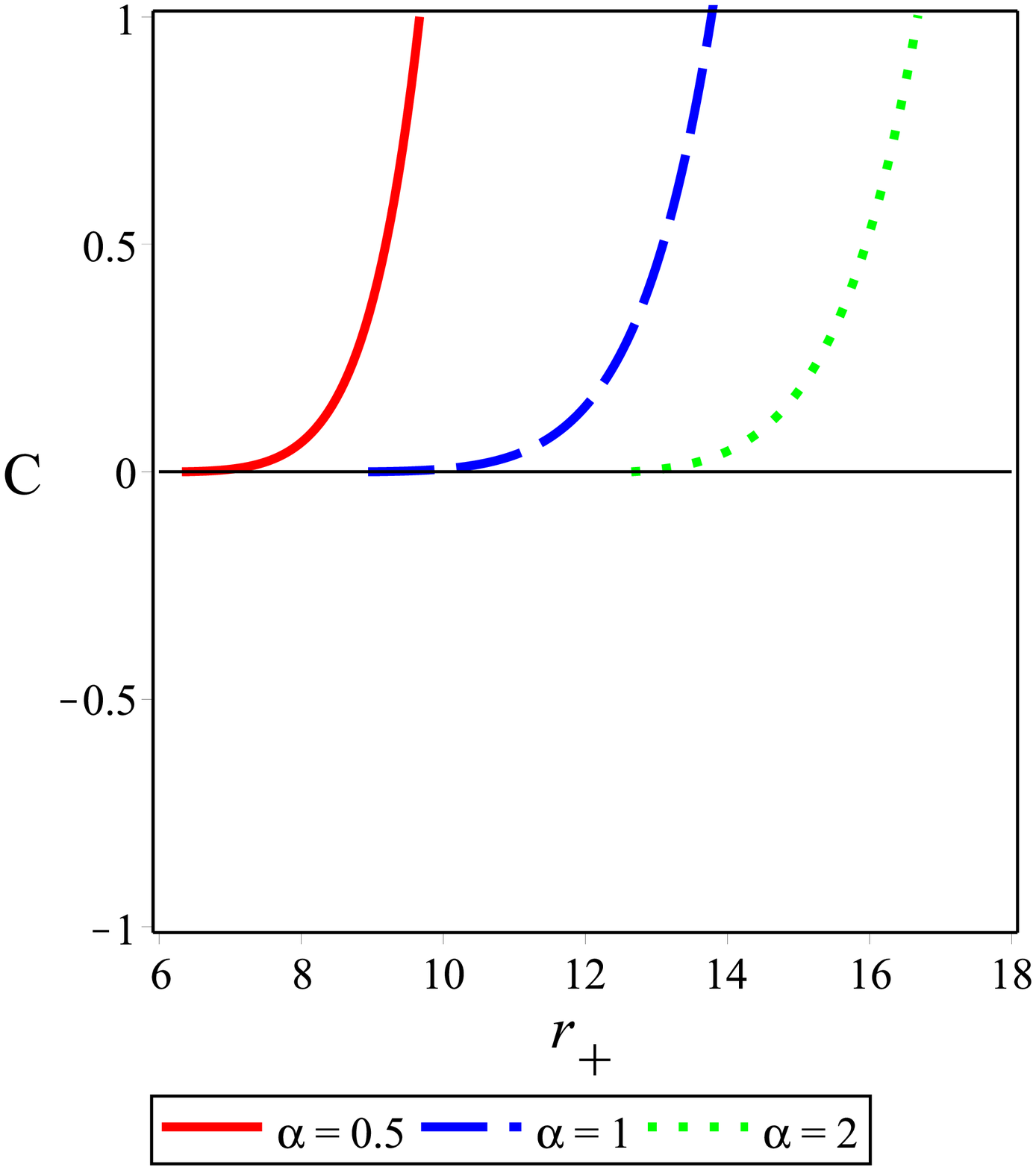}\label{C72}}\hspace*{.5cm}
\subfigure[ $10^{-14}C-r_+$ (red), $10^{-16}C-r_+$ (blue) and
$10^{-20}C-r_+$ (green) for $n=8$ and $x=2$.
    ]{\includegraphics[scale=0.25]{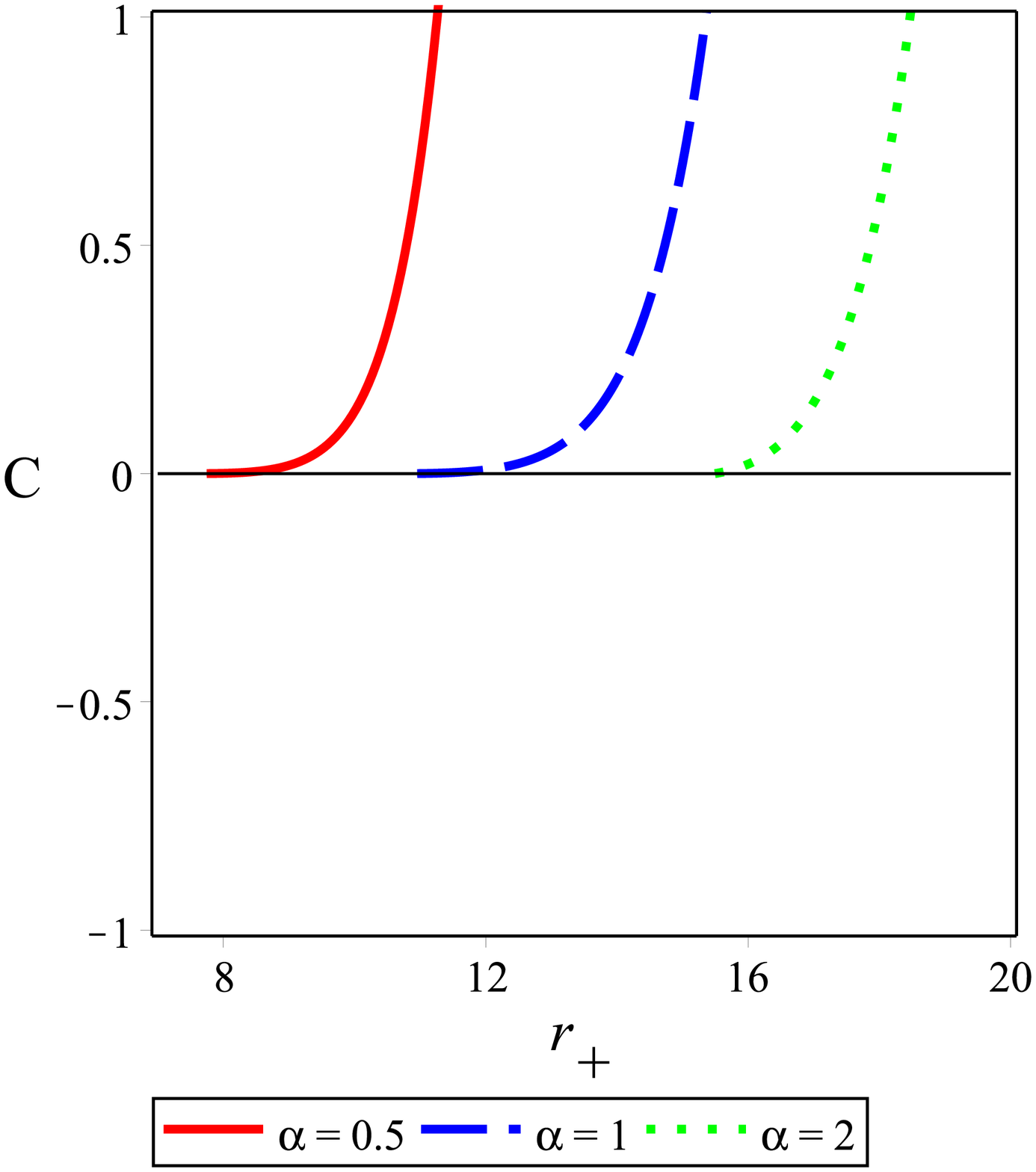}\label{C82}}
\caption{Behavior of the heat capacity versus $r_{+} $ for $G=1$ and for "$+$%
" sign branch.}
\label{Figheat1}
\end{figure}

Inserting Eqs. (\ref{mass:2}) and (\ref{Temp:2}) into Eq. (\ref{entropy}),
one can calculate the entropy as
\begin{equation}
S=\frac{1}{G\alpha}\left[ x^{x+1}r_{+}^{(n-3)x}\sqrt{\alpha^{2}(2\hat{\alpha}%
)^{x-1}} \left( r_{+}^{2x}-\frac{(n-1)\hat{\alpha}^{x}(2x)^{2(x-1)}}{%
(n-3)2^{x-2}}\right) \right] .
\end{equation}

The important quantity to study the local stability of the system is heat
capacity which is determined as
\begin{equation}  \label{Heat2}
C=\frac{(n-1)\sqrt{\tilde\alpha^{x+1}}(nx^{2}
r_{+}^{2x}-(n-4)\tilde\alpha^{x})(\tilde%
\alpha^{x}-x^{2}r_{+}^{2x})r_{+}^{(n-2)x} }{\tilde\alpha Gr_{+}^{x} \left[%
(n-8)x^{2} r_{+}^{2x}+(n-4)\tilde\alpha^{x}\right]} .
\end{equation}

The behavior of the heat capacity as a function of $r_{+}$ is depicted in
Fig. \ref{Figheat1}. For up panels of this figure, we fixed the parameter $%
\alpha$ and investigated the effect of dynamic exponent $x$ on the heat
capacity. It can be seen that for all values of $n$ and $x$ the heat
capacity is positive, meaning that the black hole is thermally stable all
the time. For down panels of this figure, we fixed the dynamic exponent $x$
and examined the influence of $\alpha$ on the heat capacity. Looking at
these graphs, one can find that the effect of parameter $\alpha$ on the heat
capacity is similar to that of the dynamic exponent $x$. So, a similar
discussion can be employed for this case.


\begin{figure}[h!]
\subfigure[ $10^{-1}F-r_+$ (red), $10^{-6}F-r_+$ (blue) and
$10^{-14}F-r_+$ (green) for $n=6$  and
$\alpha=0.5$.]{\includegraphics[scale=0.25]{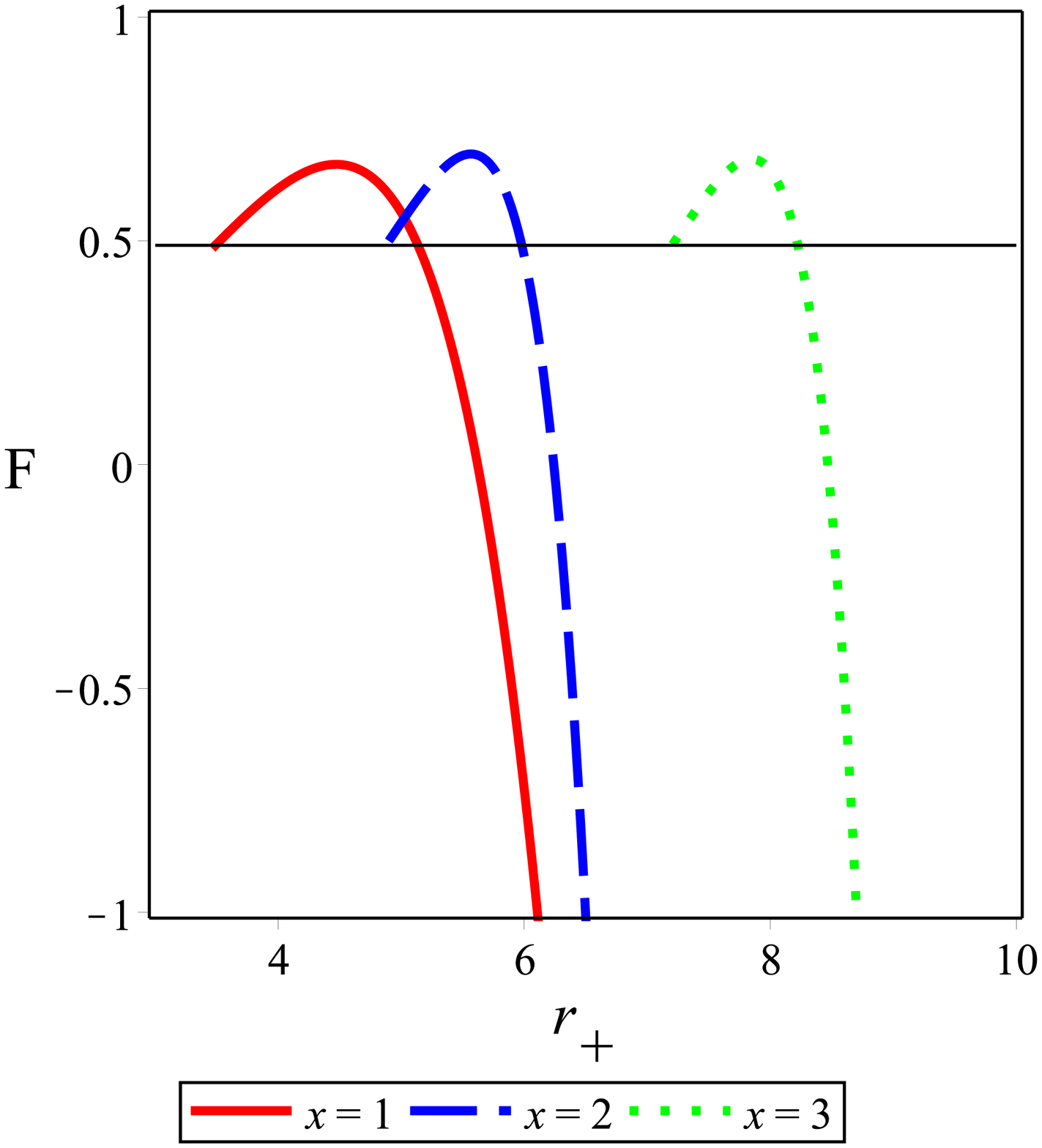}\label{F61}} \hspace*{.5cm}
\subfigure[ $10^{-3}F-r_+$ (red), $10^{-10}F-r_+$
(blue) and $10^{-19}F-r_+$ (green) for $n=7$  and  $\alpha=0.5$.
    ]{\includegraphics[scale=0.25]{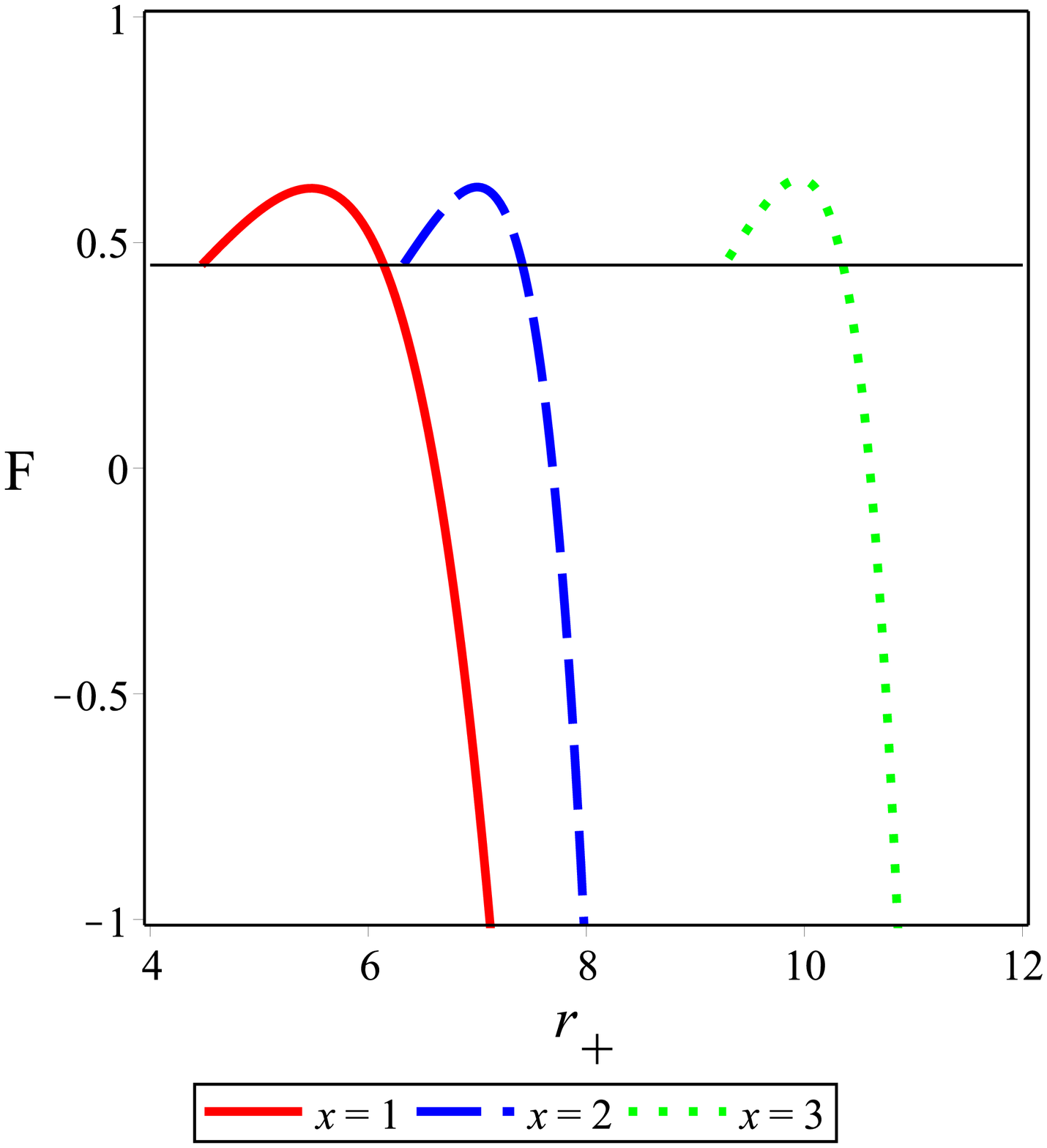}\label{F71}}\hspace*{.5cm}
\subfigure[ $10^{-4}F-r_+$ (red), $10^{-13}F-r_+$ (blue) and
$10^{-24}F-r_+$ (green) for $n=8$  and $\alpha=0.5$.
    ]{\includegraphics[scale=0.25]{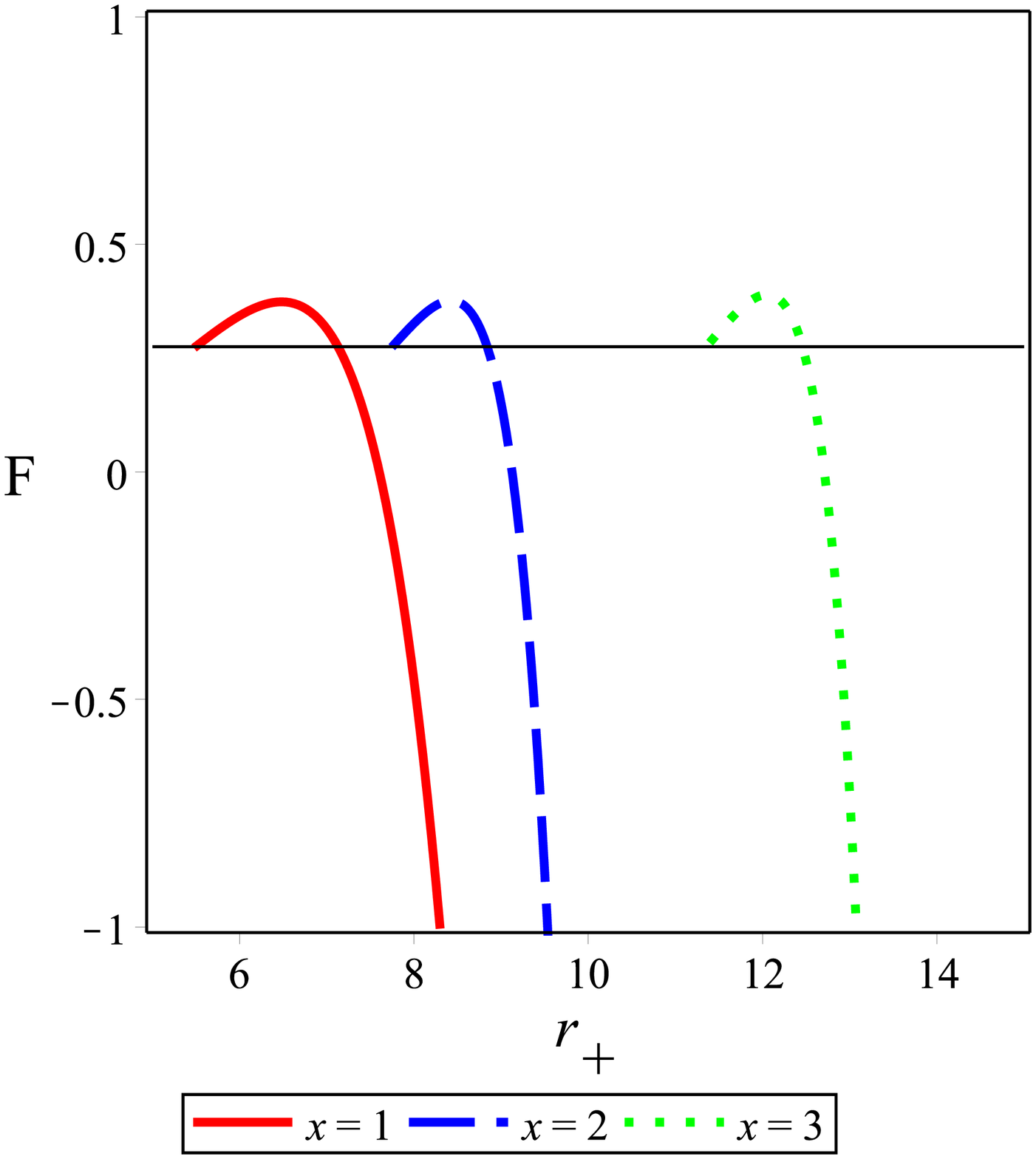}\label{F81}}\newline
\subfigure[ $10^{-6}F-r_+$ (red), $10^{-8}F-r_+$ (blue) and
$10^{-10}F-r_+$ (green) for $n=6$  and
$x=2$.]{\includegraphics[scale=0.25]{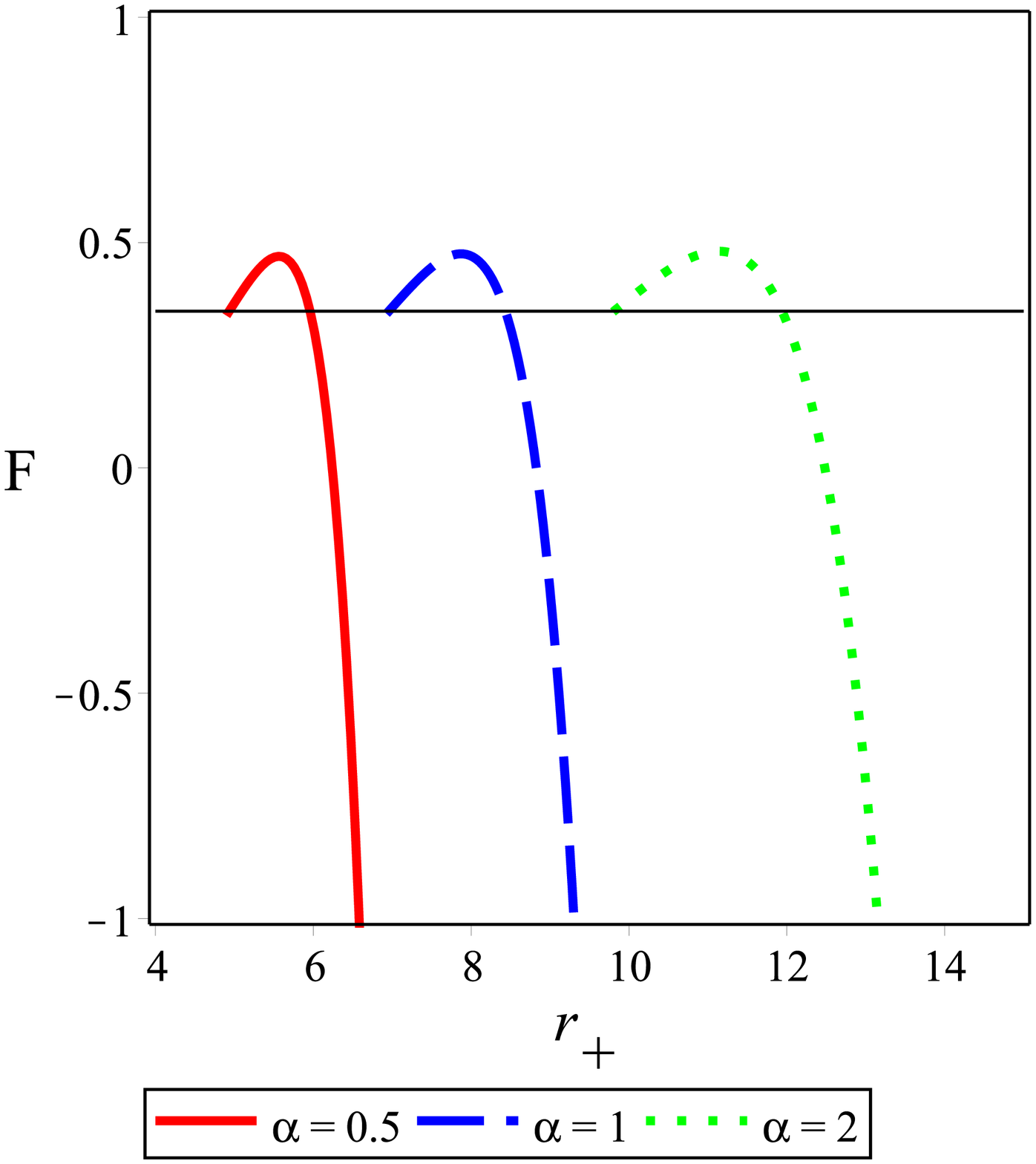}\label{F62}} \hspace*{.5cm}
\subfigure[ $10^{-10}F-r_+$ (red), $10^{-12}F-r_+$
(blue) and $10^{-14}F-r_+$ (green) for $n=7$  and  $x=2$.
    ]{\includegraphics[scale=0.25]{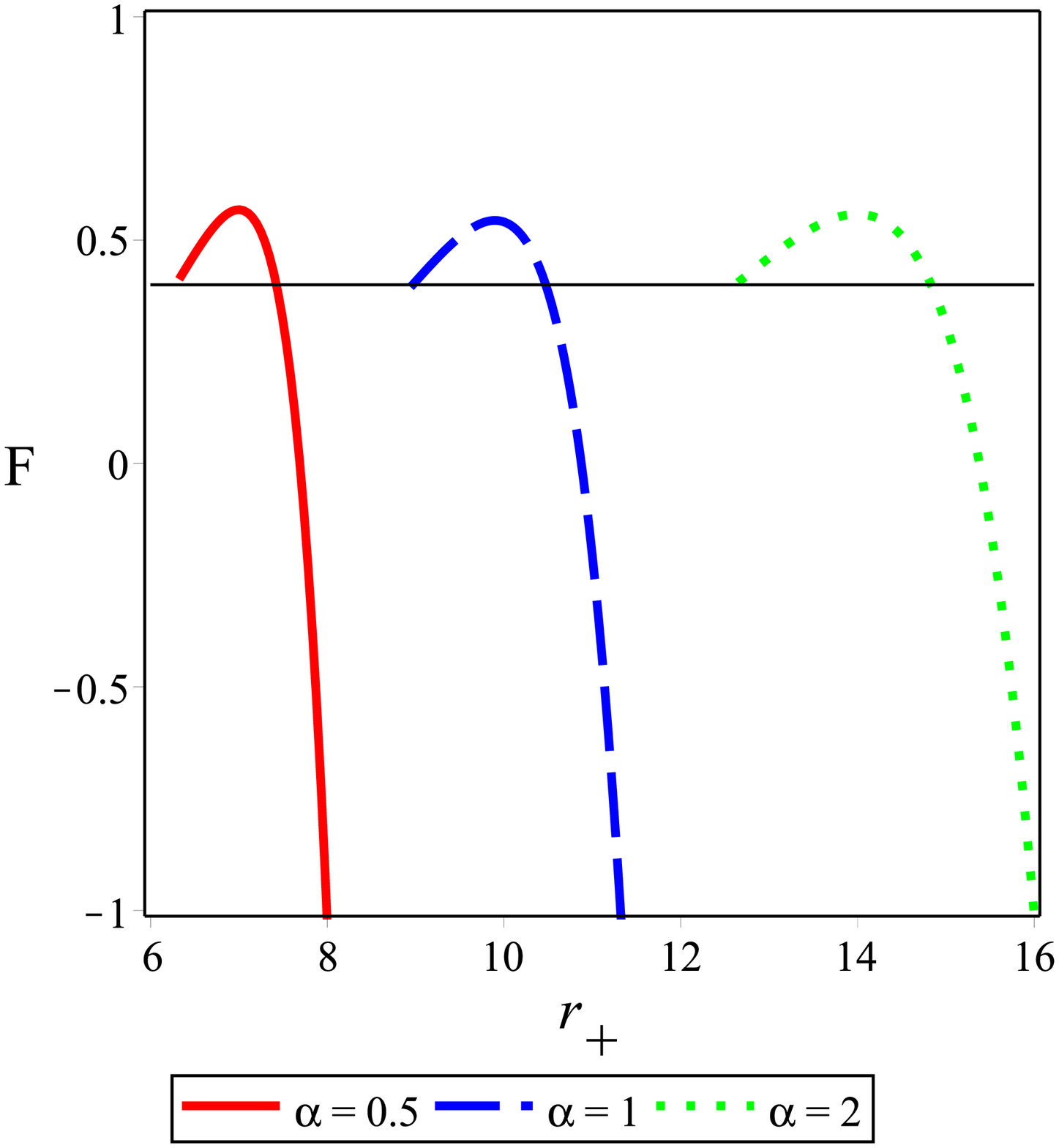}\label{F72}}\hspace*{.5cm}
\subfigure[ $10^{-13}F-r_+$ (red), $10^{-15}F-r_+$ (blue) and
$10^{-18}F-r_+$ (green) for $n=8$  and $x=2$.
    ]{\includegraphics[scale=0.25]{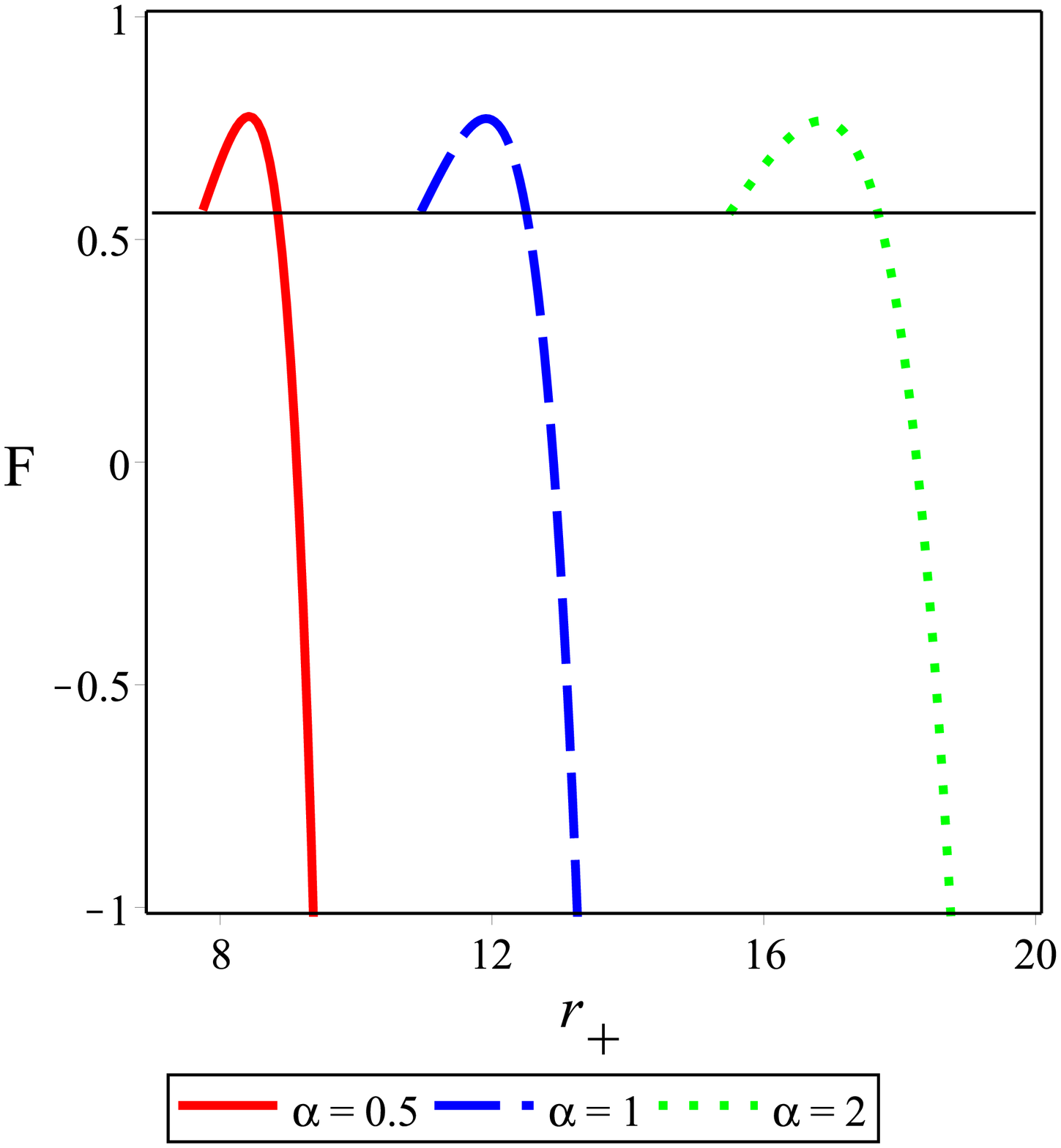}\label{F82}}
\caption{Behavior of the free energy versus $r_{+} $ for $G=1$ and for "$+$"
sign branch.}
\label{FigGibbs1}
\end{figure}

In Fig. \ref{FigGibbs1}, we studied the behavior of the free energy under
variation of the dynamic exponent $x$ (see up graphs) and variation of the
Gauss-Bonnet coefficient $\alpha$ (see down graphs). One can see that for
arbitrary dimensions, there are two sets of black holes with small and large
horizon radii. According to this figure, small black holes are unstable
whereas large black holes enjoy global stability. Besides, the stability
region changes as both parameters $x$ and $\alpha$ vary.

\section{Conclusion}\label{five}

In this paper, we proposed a noteworthy geometry in which the spatial
coordinates scale anisotropically i.e. $(t,r,\theta_{i})\,\to
(\lambda^{z}t,\lambda^{-1}r,\lambda^{x_i}\,\theta_{i})$.
AdS spacetime is one of the trivial subclasses of the mentioned
unusual geometry for $z=x_{i}=1$, while it cannot be a nontrivial
solution of the pure Einstein gravity without matter field.
However, the higher derivative curvature terms could play the role
of the desired matter fields, and therefore, we investigated
whether Gauss-Bonnet gravity could admit the proposed geometry as
a nontrivial solution in the vacuum and understood that such a
solution exists in this kind of gravity under certain
circumstances (see Eq.\eqref{Coef1}) and provided that
$x_{i}$'s$=x$. In addition, we found that there is an exact vacuum
solution for $k=\pm\,1$, independent of the value of the dynamical
exponent $z$, which is a black hole solution for pseudo-hyperbolic
horizon structure $k=-1$ while the pseudo-spherically symmetric
boundary ($k=1$) results in a naked singularity. It is also
notable that according to our calculations, $5-$dimensional
spacetime is a special case for which the critical exponent $x$
can only take the value of the unit ($x=1$). Whereas there is no
restriction on the value of the parameter $x$ in higher dimensions
($n>4$).

Besides, we looked for a general solution of our suggested metric
in Einstein-Gauss-Bonnet gravity and found that there is an exact
solution provided the parameters of the model (cosmological
constant as well as Gauss-Bonnet coefficient) are chosen suitably
and the dynamical exponents $x$ and $z$ have the same values.
Given the mandatory constraints on the model, the obtained
solution is equivalent to the AdS solution with $\hat{\alpha}=-\frac{n(n-1)}{%
8\Lambda}$ and considering the fact that the cosmological constant in our
model for the case of $x=1$ is half of that in AdS spacetime. Since the
obtained solution has two branches with $"+"$ and $"-"$ signs, we
investigated the properties of each branch of the solution, separately. We
found that the metric function with the $"-"$ sign can be interpreted as a
black hole with just one horizon for three types of geometric structures ($%
k=0\,,\pm 1$). Moreover, studying the role of varying $x$ on the properties
of $"-"$, one finds that the solution with $k=1$ is much more sensitive to
the change of $x$ than the solutions with $k=0$ and $k=-1$.

We also examined the thermodynamic properties of the black hole solutions
for the branch with $"-"$ sign, and then, we studied the stability of the
obtained solution for each geometrical structure of the boundary (boundary
with $r=cte$ and $t=cte$), separately. The obtained results are

\begin{itemize}
\item Concerning thermodynamic behavior of the black hole with $k=0$
boundary, we found that this solution enjoys local stability due to the
positivity of the temperature and heat capacity for all values of the model
parameters. Furthermore, strictly decreasing the behavior of the free energy
function assures us of the global stability of the solution.

\item For the black holes with pseudo-hyperbolic horizon structure ($k=-1$),
our calculations indicated that the solutions are not always stable. We
calculated the smallest horizon radius of the stable black hole (${r_{+}}_{%
{\tiny {\text{min}}}}$) and investigated the thermodynamic behavior of the
stable system. Our investigation showed that the temperature of the stable
black hole starts from $T_{{\tiny {\text{min}}}}$ at ${r_{+}}_{{\tiny {\text{%
min}}}}$ and monotonically goes to infinity as $r_{+}\rightarrow\infty$.
Also, the free energy always starts from zero, reaches a positive maximum
value $F_{{\tiny \text{max}}}$, and then goes to negative infinity as $r_+
\rightarrow\,\infty$.

\item Regarding the pseudo-spherically symmetric horizon ($k=+1$), it was
found that although the temperature is always positive for each value of
model parameters, the heat capacity experiences one divergency at $r_{{\tiny
{\text{div}}}}$ and meets positive values just for a range of horizon radii
which means that the solution is not always stable. We understood that if
the radius of the event horizon is larger than $r_{{\tiny {\text{div}}}}$,
the black hole is thermally stable and enjoys some necessary criteria for
viable solutions. Besides, the free energy function always meets a positive
maximum value at some $r_+$ and tends to negative infinity when $r_+
\rightarrow\,\infty$.
\end{itemize}

On the other hand, regarding the branch of metric function with "$+$" sign,
we understood that black hole solutions only exist for pseudo-hyperbolic
horizon ($k=-1$) and other curvature result in a naked singularity. It was
also found that, depending on the metric parameters, this solution could
represent a black hole with two horizons, an extreme black hole with a
degenerate root or a naked singularity. Next, we calculated the
thermodynamic quantities in arbitrary dimensions and focus on the thermal
behavior of the system. As expected, we found that black holes with large
horizon radius is thermally stable.

Regarding this special anisotropic scaling spacetime, it will be interesting
to study dynamic stability and related quasinormal modes. Moreover, it is
worth to investigate modified dispersion relations in this geometry.
Besides, one can generalize the obtained solutions in the presence of matter
field. Furthermore, analyzing the effects of higher order curvature terms
(such as third order Lovelock gravity) will be interesting. Also, according
to obtained results, we found that the geometrical properties of the
positive branch with $k=-1$ are more or less similar to the asymptotically
flat charged black hole with a spherical horizon. So, it is worth
investigating possible difference (or more similarities) as well as the
causality of the solution according to the Penrose diagram since the
solution enjoys timelike singularity. These topics should be addressed in
independent works.


\section*{Acknowledgments}

S. Mahmoudi and Kh. Jafarzade are grateful to the Iran Science Elites
Federation for the financial support.


\begin{thebibliography}{99}
\bibitem{Witten:1998qj} E.~Witten,
Adv. Theor. Math. Phys. \textbf{2}, 253 (1998)

\bibitem{Gubser:1998bc} S.~S.~Gubser, I.~R.~Klebanov and A.~M.~Polyakov,
Phys. Lett. B \textbf{428}, 105 (1998) 

\bibitem{Maldacena:1997re} J.~M.~Maldacena,
Adv. Theor. Math. Phys. \textbf{2}, 231 (1998) 

\bibitem{Hartnoll:2009sz} S.~A.~Hartnoll,
Class. Quant. Grav. \textbf{26}, 224002 (2009)

\bibitem{Kovtun:2004de} P.~Kovtun, D.~T.~Son and A.~O.~Starinets,
Phys. Rev. Lett. \textbf{94}, 111601 (2005)

\bibitem{Maldacena:2008wh} J.~Maldacena, D.~Martelli and Y.~Tachikawa,
JHEP \textbf{10}, 072 (2008) 

\bibitem{Herzog:2008wg} C.~P.~Herzog, M.~Rangamani and S.~F.~Ross,
JHEP \textbf{11}, 080 (2008) 

\bibitem{Herzog:2009xv} C.~P.~Herzog,
J. Phys. A \textbf{42}, 343001 (2009) 

\bibitem{Mehen:1999nd} T.~Mehen, I.~W.~Stewart and M.~B.~Wise,
Phys. Lett. B \textbf{474}, 145 (2000) 

\bibitem{Son:2008ye} D.~T.~Son,
Phys. Rev. D \textbf{78}, 046003 (2008) 

\bibitem{Balasubramanian:2008dm} K.~Balasubramanian and J.~McGreevy,
Phys. Rev. Lett. \textbf{101}, 061601 (2008)

\bibitem{Kachru:2008yh} S.~Kachru, X.~Liu and M.~Mulligan,
Phys. Rev. D \textbf{78}, 106005 (2008) 

\bibitem{Cardy:2002} J. Cardy, Scaling and Renormalization in Statistical
Physics (Cambridge University Press, Cambridge, 2002)

\bibitem{Taylor:2008tg} M.~Taylor, 
[arXiv:0812.0530 [hep-th]]

\bibitem{Dehghani:2010kd} M.~H.~Dehghani and R.~B.~Mann,
JHEP \textbf{07}, 019 (2010) 

\bibitem{Dehghani:2010gn} M.~H.~Dehghani and R.~B.~Mann,
Phys. Rev. D \textbf{82}, 064019 (2010) 

\bibitem{Brenna:2011gp} W.~G.~Brenna, M.~H.~Dehghani and R.~B.~Mann,
Phys. Rev. D \textbf{84}, 024012 (2011) 

\bibitem{Ghanaatian:2014bpa} M.~Ghanaatian, A.~Bazrafshan and W.~G.~Brenna,
Phys. Rev. D \textbf{89}, 124012 (2014) 

\bibitem{Maeda:2011jj} H.~Maeda and G.~Giribet,
JHEP \textbf{11}, 015 (2011) 

\bibitem{Lee:2010iu} J.~Lee, T.~H.~Lee and P.~Oh,
Phys. Lett. B \textbf{701}, 393 (2011) 

\bibitem{Alvarez:2014pra} A.~Alvarez, E.~Ay\'on-Beato, H.~A.~Gonz\'alez and
M.~Hassa\"\i{}ne,
JHEP \textbf{06}, 041 (2014) 

\bibitem{Lovelock:1971} D. Lovelock, J. Math. Phys. \textbf{12}, 498 (1971)

\bibitem{stringth1} B. Zwiebach, Phys. Lett. B \textbf{156}, 315 (1985)

\bibitem{stringth2} D. J. Gross and E. Witten, Nucl. Phys. B \textbf{277}, 1
(1986)

\bibitem{stringth3} D. J. Gross and J. H. Sloan, Nucl. Phys. B \textbf{291},
41 (1987)

\bibitem{stringth4} R. R. Metsaev and A. A. Tseytlin, Phys. Lett. B \textbf{%
191}, 354 (1987)

\bibitem{stringth5} R. R. Metsaev and A. A. Tseytlin, Nucl. Phys. B \textbf{%
293}, 385 (1987)

\bibitem{GB1} C. Lanczos, Ann. Math. \textbf{39} , 842 (1938)

\bibitem{Boulware:1985wk} D.~G.~Boulware and S.~Deser,
Phys. Rev. Lett. \textbf{55}, 2656 (1985) 

\bibitem{Zumino:1985dp} B.~Zumino,
Phys. Rept. \textbf{137}, 109 (1986) 

\bibitem{GB3} Y. M. Cho, I. P. Neupane and P. S. Wesson, Nucl. Phys. B
\textbf{621} , 388 (2002) 

\bibitem{GB4} R. G. Cai, Phys. Rev. D \textbf{65} , 084014 (2002)

\bibitem{ghost1} R. Myers, Nucl. Phys. B \textbf{289}, 701 (1987)

\bibitem{ghost2} C. Callan, R. Myers and M. Perry, Nucl. Phys. B \textbf{311}%
, 673 (1989) 
\bibitem{Nojiri:2005vv}
S.~Nojiri, S.~D.~Odintsov and M.~Sasaki,
Phys. Rev. D \textbf{71}, 123509 (2005)
\bibitem{Nojiri:2010wj}
S.~Nojiri and S.~D.~Odintsov,
Phys. Rept. \textbf{505}, 59-144 (2011)
doi:10.1016/j.physrep.2011.04.001
[arXiv:1011.0544 [gr-qc]].




\bibitem{Nojiri:2000gv} S.~Nojiri and S.~D.~Odintsov,
JHEP \textbf{07}, 049 (2000) 

\bibitem{blackGB1} T. Kobayashi and T. Tanaka, Phys. Rev. D \textbf{71},
084005 (2005)

\bibitem{blackGB2} C. Bogdanos, C. Charmousis, B. Gouteraux and R. Zegers,
JHEP \textbf{10}, 037 (2009)

\bibitem{blackGB3} Y. Brihaye, T. Delsate and E. Radu, JHEP \textbf{07}, 022
(2010) 

\bibitem{blackGB4} G. Giribet, J. Oliva and R. Troncoso, JHEP \textbf{05},
007 (2006)

\bibitem{blackGB5} A. Giacomini, J. Oliva and A. Vera, Phys. Rev. D \textbf{%
91}, 104033 (2015)

\bibitem{holog1} R. A. Konoplya and A. Zhidenko, JHEP \textbf{09}, 139
(2017) 

\bibitem{holog2} S. Grozdanov and A. O. Starinets, JHEP \textbf{03}, 166
(2017) 

\bibitem{holog31} Y. Sun, Hao Xu, L. Zhao, JHEP \textbf{09}, 060 (2016)

\bibitem{holog32} Y. Z. Li, S. F. Wu and G. H. Yang, Phys. Rev. D \textbf{88}%
, 086006 (2013)

\bibitem{holog33} X. X. Zeng, X. M. Liu and W. B. Liu, JHEP \textbf{03}, 031
(2014) 

\bibitem{holog34} S. J. Zhang, B. Wang, E. Abdalla and E. Papantonopoulos,
Phys. Rev. D \textbf{91}, 106010 (2015)

\bibitem{holog4} R. Gregory, S. Kanno and J. Soda, JHEP \textbf{10}, 010
(2009) 

\bibitem{holog5} L. Barclay, R. Gregory, S. Kanno and P. Sutcliffe, JHEP
\textbf{12}, 029 (2010)

\bibitem{holog6} A. Sheykhi, H. R. Salahi and A. Montakhab, JHEP \textbf{04}%
, 058 (2016)


\bibitem{Con1} P.~Ghaemi, A.~Vishwanath and T.~Senthil, Phys. Rev. \textbf{B72}, 024420 (2005)

\bibitem{Con2} S.~Sachdev and T.~Senthil, Ann. Phys. \textbf{251}, 76 (1996)

\bibitem{Con3} K. Yang, Phys. Rev. Lett. \textbf{93}, 066401 (2004)

\bibitem{Con4} M.~Freedman, C.~Nayak, and K.~Shtengel, Phys. Rev. Lett.
\textbf{94}, 147205 (2005) 

\bibitem{Mann:1a} R. B. Mann et al. Sci. Rep. \textbf{11}, 7474 (2021)%


\bibitem{Kiefer:1a} C. Kiefer and M. Kramer, Phys. Rev. Lett. \textbf{108},
021301 (2012)

\bibitem{Bini:2a} D. Bini et al. Phys. Rev. D \textbf{87}, 104008 (2013)

\bibitem{Bucher:2ab} M. Bucher, Int. J. Mod. Phys. D \textbf{24}, 1530004
(2015) 

\bibitem{Mateos:2ab} D. Mateos and D. Trancanelli, JHEP \textbf{07}, 054
(2011) 

\bibitem{Florkowski:2ab} W. Florkowski, Phys. Lett. B \textbf{668}, 32
(2008)

\bibitem{Horava:2009uw} P.~Horava,
Phys. Rev. D \textbf{79}, 084008 (2009) 


\bibitem{Dotti:2007az} G.~Dotti, J.~Oliva and R.~Troncoso,
Phys. Rev. D \textbf{76}, 064038 (2007) 


\bibitem{Bardeen:1973gs}
J.~M.~Bardeen, B.~Carter and S.~W.~Hawking,
Commun. Math. Phys. \textbf{31}, 161-170 (1973)

\bibitem{Chatterjee:2012zh}
S.~Chatterjee, D.~A.~Easson and M.~Parikh,
Class. Quant. Grav. \textbf{30}, 235031 (2013)

\bibitem{Kontou:2020bta}
E.~A.~Kontou and K.~Sanders,
Class. Quant. Grav. \textbf{37}, no.19, 193001 (2020)




\end{thebibliography}
\end{document}